\documentclass[twocolumn]{aastex62}
\usepackage{xcolor}

\usepackage{natbib}

\shorttitle{Origin and evolution of LPCs}
\shortauthors{Vokrouhlick\'y et al.}

\begin{document}

\title{Origin and evolution of long-period comets}



\author{David Vokrouhlick\'y}
\affil{Institute of Astronomy, Charles University,
       V Hole\v{s}ovi\v{c}k\'ach 2, CZ--180 00 Prague 8,
       Czech Republic}

\author{David Nesvorn\'y}
\affil{Department of Space Studies, Southwest Research Institute,
       1050 Walnut St., Suite 300, Boulder, CO, 80302, USA}

\author{Luke Dones}
\affil{Department of Space Studies, Southwest Research Institute,
       1050 Walnut St., Suite 300, Boulder, CO, 80302, USA}


\begin{abstract}
 We develop an evolutionary model of the long-period comet (LPC)
 population, starting from their birthplace in a massive
 trans-Neptunian disk that was dispersed by migrating giant
 planets. Most comets that remain bound to the Solar system are stored
 in the Oort cloud. Galactic tides and passing stars make some of these
 bodies evolve into observable comets in the inner Solar system. Our
 approach models each step in a full-fledged numerical
 framework. Subsequent analysis consists of applying plausible fading
 models and computing the original orbits to compare with observations.
 Our results match the observed semimajor axis distribution of LPCs
 when Whipple's power-law fading scheme with an exponent $\kappa=
 0.6^{+0.1}_{-0.2}$ is adopted. The cumulative perihelion ($q$)
 distribution is fit well by a linear increase plus a weak quadratic
 term. Beyond $q = 15$~au, however, the population increases steeply
 and the isotropy of LPC orbital planes breaks. We find tentative
 evidence from the perihelion distribution of LPCs that the returning
 comets are depleted in supervolatiles and become active due to water
 ice sublimation for $q\leq 3$ au. Using an independent calibration
 of the population of the initial disk, our predicted LPC flux is
 smaller than observations suggest by a factor of $\simeq 2$.
 Current data only characterize comets from the outer Oort cloud
 (semimajor axes $\gtrsim 10^4$~au). A true boost in understanding the
 Oort cloud's structure should result from future surveys when they
 detect LPCs with perihelia beyond $15$~au. Our results provide
 observational predictions of what can be expected from these new
 data.
\end{abstract}



\section{Introduction}
Comets are primitive bodies born mostly in a massive trans-Neptunian
disk, though some might have formed in the region in between giant
planets too. They share a birthplace with several other populations of
small bodies in the outer Solar system, such as Jupiter and Neptune
Trojans, the irregular satellites of giant planets, the resonant and
hot components of the Kuiper belt, and objects in the scattering
disk. Out of all these categories of small bodies, comets underwent
the most spectacular orbital evolution before being observed. Except
for those in the Jupiter family, comets were scattered by the giant
planets to the very outskirts of the Solar system to form a storage
zone called the Oort cloud. There, barely gravitationally bound to the
Sun, comets wait eons for their chance to return to the inner regions
of the Solar system. Assisted by galactic tides and tugs from passing
stars, they eventually set on their journeys.  They plunge into the
planetary zone on highly eccentric orbits before disappearing forever
\citep[e.g.][]{dones2004}. Obviously, their activity -- namely, the
production of gas and dust comae as they become heated by solar
radiation when they get close enough to the Sun -- makes them
classified as comets in the first place and constitutes the glory 
of their deadly run.

Cometary precursors in the Oort cloud cannot be observed in situ. This
holds even for the largest expected members in this population, which
may be Pluto-sized, or even larger. Therefore, unraveling properties
of the Oort cloud remains one of the great challenges in planetary
science. They can only be inferred thus far from observations of
comets that once visited the Oort cloud region. Halley-type comets
(HTC) are less useful in this respect. This is because before being
observed, HTCs underwent significant orbital evolution after leaving
their source zone. Therefore, the long-period comets (LPCs) are a
better tracer population of the Oort cloud. Using the commonly adopted
definition, we define LPCs as comets with orbital periods longer than
$200$~yr (thus heliocentric semimajor axis $a\gtrsim 35$~au).
However, most LPCs reside on much
more extreme orbits having $a$ equal to thousands or even tens of
thousands of au. The equivalent orbital periods are as large as
several million years. With these orbital parameters, LPCs can tell us
a great deal about the Oort cloud architecture.

The fundamental facts about LPC orbits have been pinned down already
by \citet{oort1950}: (i) a preponderance of comets on nearly parabolic
orbits, constituting what is now called the Oort peak, with the
implication of strong fading during subsequent returns (see
Section~\ref{fad}), (ii) near isotropy of the orbital planes in space,
and (iii) nearly equal numbers of LPCs in equal bins of perihelia
for $q < 1.5$~au. It is somewhat surprising how little has
been added to this broad picture on the observational side over the
past decades, especially if compared with the vast increase of data
about other populations of small bodies in the Solar system. The
additions include (i) a more complete characterization of the
returning population of LPCs on orbits more strongly bound to the Sun,
and (ii) extension of the data set to larger perihelia. The paucity of
new data is due, in part, because, until the late 1990s, only about a
dozen or fewer new LPCs were discovered annually, many by amateurs,
rather than by well-characterized surveys
(\url{http://comethunter.de/}). The situation has
improved in the past two decades, but a significant boost of new LPC
discoveries by surveys is still in the future.

The theory side of LPC studies has evolved somewhat more. It has been
understood that the inner edge of the Oort peak at about $10000$~au
is simply an apparent structure due to a bias
related to observing only comets with small perihelion distances
\citep[e.g.,][]{hills1981}. The inner Oort cloud is expected to extend
to $\approx$~3000~au from the Sun \citep[e.g.,][]{Duncan1987AJ}, but comets
from the inner cloud should only reach the inner Solar system during
rare comet showers \citep[e.g.,][]{hetal1987,h1990}.  The role of the Sun's likely
birth cluster, the Sun's migration in the Galaxy, and planetary
migration were all investigated.  The dynamics of bodies stored in the
Oort cloud was also understood by analyzing the effects of galactic
tides and stellar short-range perturbations. Finally, other studies
shed a detailed light on the transfer dynamics of comets into the
heliocentric zone where they become observable.  Reviews may be found
in \citet{dones2004}, \citet{rickman2010} and \citet{detal2015}.

In spite of all these improvements, and partially because of lack of
data, fewer studies were devoted to a direct comparison of theoretical
predictions with LPC observations. An outstanding achievement in this
respect was obtained by \citet{WT99}, who compared the available data
to the state of the art in modeling of LPC dynamics.  Still, this work
adopted a number of simplifications. For instance, all available data
were compressed into three measures which the authors confronted with
model predictions: (i) the number of comets in the Oort peak vs. all
LPCs, (ii) the number of comets in the small-semimajor axis tail
(34.5~au$\lesssim a\lesssim 69$~au) vs. all LPCs, and (iii) the
number of comets with retrograde orbits vs.\ all LPCs. These data
constrain the model in its important aspects, yet they remain rather
coarse. The numerical model used in \citet{WT99} was obviously
restricted by computer capabilities at that time, but it also
neglected some important effects. For instance, prevailing opinion in
the 1990s highlighted the effects of galactic tides over the
perturbations due to passing stars. However, further analyses found
about equal importance -- or even a synergistic role -- of both
effects \citep[e.g.,][]{retal2008}.

Our goal in this work is to extend the effort of \citet{WT99} in both
aspects, namely orbital data and numerical model. As for the data
side, we have now more complete information. Significant improvements
especially concern the class of LPCs on near-parabolic orbits
(Sec.~\ref{war}). There have been new estimates of the annual flux of
LPCs, though uncertainties still remain about the sizes of cometary
nuclei \citep[e.g.,][]{francis2005, bm2013}. There has been again more
improvement on the modeling side. Most importantly, today's computer
capabilities allow us to propagate (i) the orbits of millions of test
particles from their ultimate birthplaces to the moments they become
observable as comets some $4.5$~Gyr later, and (ii) use a single
framework of a full-fledged N-body integrator (without switching
between a secular approximation and an N-body calculation). A unique
aspect of our approach consists of using initial orbital data for
comets that reflect their true birth zone, which has been calibrated
by other, independent applications of the model. Finally, our work
complements the model presented in \citet{nesvornySPC}, where the
origin and dynamical evolution of short-period comets was analyzed and
confronted with observations. Therefore, it is for the first time --
to our knowledge -- that the same model is used to explain the
properties of all comets.

In Section~2, we summarize observational data about LPCs. This has two
facets: (i) orbital architecture, principally the semimajor axis
distribution, complemented with information about perihelia and
inclinations, and (ii) the observed flux of LPCs. We focus principally
on orbits. This is because the flux information suffers uncertainty in
the magnitude-size relation of these comets. In Section~3, we present
our model. We highlight our beginning-to-end approach, following
comets from their birth environment in a dynamically cold,
trans-Neptunian disk of planetesimals to the Oort cloud and back to
the observable zone.  In Section~4, we describe results from our
simulations. First, we characterize the orbits of new and returning
comets in a chosen heliocentric target zone.  We use heliocentric
distances $r\leq 5$~au,
relevant for the population of the currently observed LPCs, and $r\leq
20$~au, in anticipation of future surveys. Next, we compare
simulations to the observations. Finally, in Section~5 we use our
model to highlight a few predictions relevant for future surveys that
should be able to detect LPCs with distant perihelia.


\section{Properties of known LPCs}\label{obs}
As we await powerful, well-characterized surveys that will provide
accurate and homogeneous information on the orbital distribution and
flux of LPCs, we are left with a sample obtained by many different
sources and different observational circumstances, often analyzed by
different computational methods. This inevitably implies biases which
cannot be entirely removed.  Cometary activity, especially at small
heliocentric distances, does not help the situation.  It not only
necessitates including complicated nongravitational effects in the
orbit determination, and thus characterization of the orbital binding
energy with which the comet approached the inner Solar system, but it
also makes it hard to determine the size of the nucleus.

With that gloomy preamble it is, however, true that tremendous steps
forward have been taken over the past decades. These efforts started
in the 1960s and resulted in the first population-wide orbital
information about LPCs in the 1970s \citep[e.g.,][]{ms1973,
  marsden1978}. Since then, Marsden and collaborators carried out
continuous improvements in orbital characterization of LPCs,
maintaining and periodically updating their catalog. The latest, 17th
edition from 2008 \citep[][MWC08]{MW08} still represents the current
state-of-the-art. In Sec.~\ref{mwc} we describe a subset of MWC08 that
will be used for comparison with our modeled LPC population.

An effort specific to LPCs on nearly parabolic orbits, roughly
speaking, those in the Oort peak with $a\gtrsim 15000$~au, has been
conducted by a group of Polish astronomers since 1970. This work
culminated with the publication of a catalog of their orbits by
\citet{ketal2014} and \citet{K2014}, later complemented by an analysis
of large-perihelion LPCs in \citet{KD2017}. A large fraction, between
20 to 50\% (depending on perihelion distance), of entries in the
catalog are comets with accurate orbits for which nongravitational
effects were included in the orbit determination from the
observations. Importantly, each orbital element, including those with
which comets approached the Solar system, is provided with a
statistical uncertainty (reflecting the specific orbital determination
accuracy).  The catalog is accompanied by a series of papers
\citep[e.g.,][]{KD2010, DK2011,KD2013,DK2015} which thoroughly
describe various aspects of the past and future motion of very
weakly-bound LPCs. Finally, this source contains comets observed through
2013, five years past the release of MWC08. In the case of comets on
nearly parabolic orbits, we thus consider the Polish catalog as a
superior source and describe its characteristics in Sec.~\ref{war}.

The orbital catalogs mentioned above do not contain information about
physical parameters of the comets (such as the absolute brightness and
size), nor do they directly describe their flux to the inner parts of
the Solar system. These data have to be inferred from other sources,
some of which are recalled in Sec.~\ref{flu}.

\subsection{Orbital characteristics of all LPCs}\label{mwc}
The MWC08 catalog contains information about the original orbits for
499 LPCs. Their orbital elements are (i) referred to the barycenter of
the Solar system, and (ii) computed from state vectors (position and
velocity) at a sufficiently large distance along the orbit prior to
each comet's passage through the planetary region (in MWC08 a distance
of 60~au is used). This definition
requires backward propagation of the osculating solution, determined
from observations at small heliocentric distances, for at least the
nominal orbit (ideally, though, also with mapping its
uncertainty). The transformation between osculating (heliocentric) and
original (barycentric) elements has the most profound effect on the
orbital semimajor axis $a$: often a formally hyperbolic heliocentric
orbit becomes elliptical. Other elements, such as perihelion distance
$q$ and inclination $i$, are less affected. Since the source of LPCs is
very distant from the inner parts of the Solar system, the barycentric
orbital elements are the most relevant for their study. As a result,
in what follows we shall always use the original orbital elements,
including the semimajor axis, in our discussion (unless specifically
mentioned otherwise).
\begin{figure}[t!]
\epsscale{1.1}
\plotone{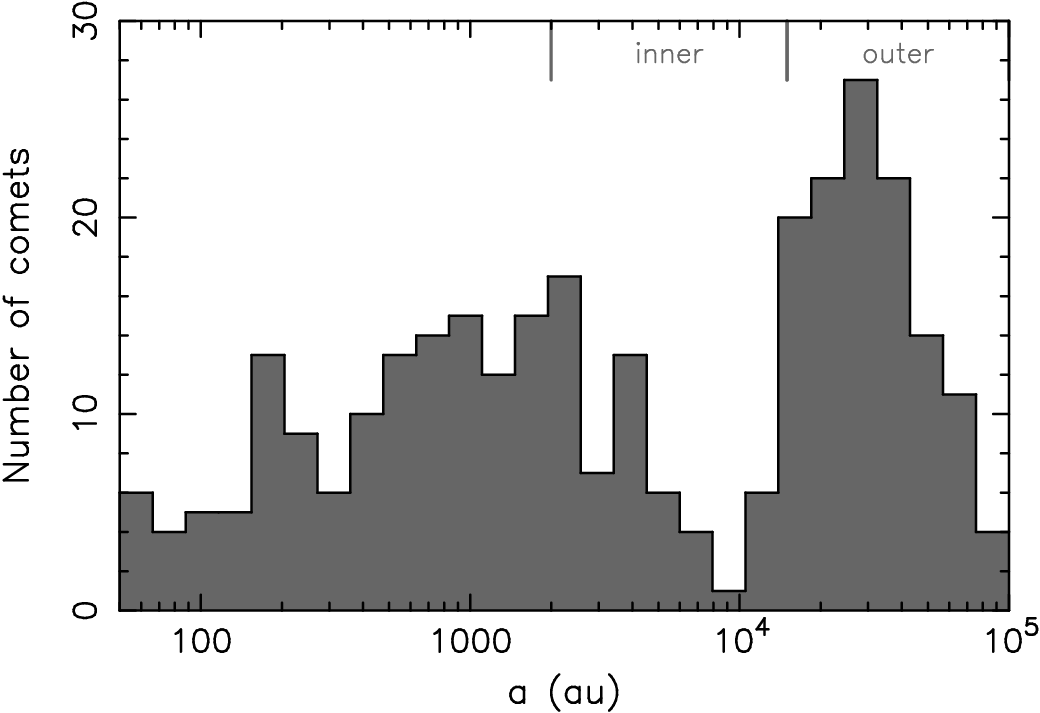}
\caption{Distribution of semimajor axes $a$ of LPCs in the MWC08
 catalog. Data for $318$ 1A and 1B orbits are used and plotted using
 equal size bins in ${\rm log}\,a$. The comets on nearly parabolic
 orbits, $a\gtrsim 15000$~au, have a source in the outer part of
 the Oort cloud. Comets having orbits with $a\lesssim 15000$~au
 generally are returning to the inner Solar system
 after they passed through the planetary zone at least once in the
 recent past.}
 \label{mw_a}
\end{figure}
\begin{figure}[t!]
\epsscale{1.1}
\plotone{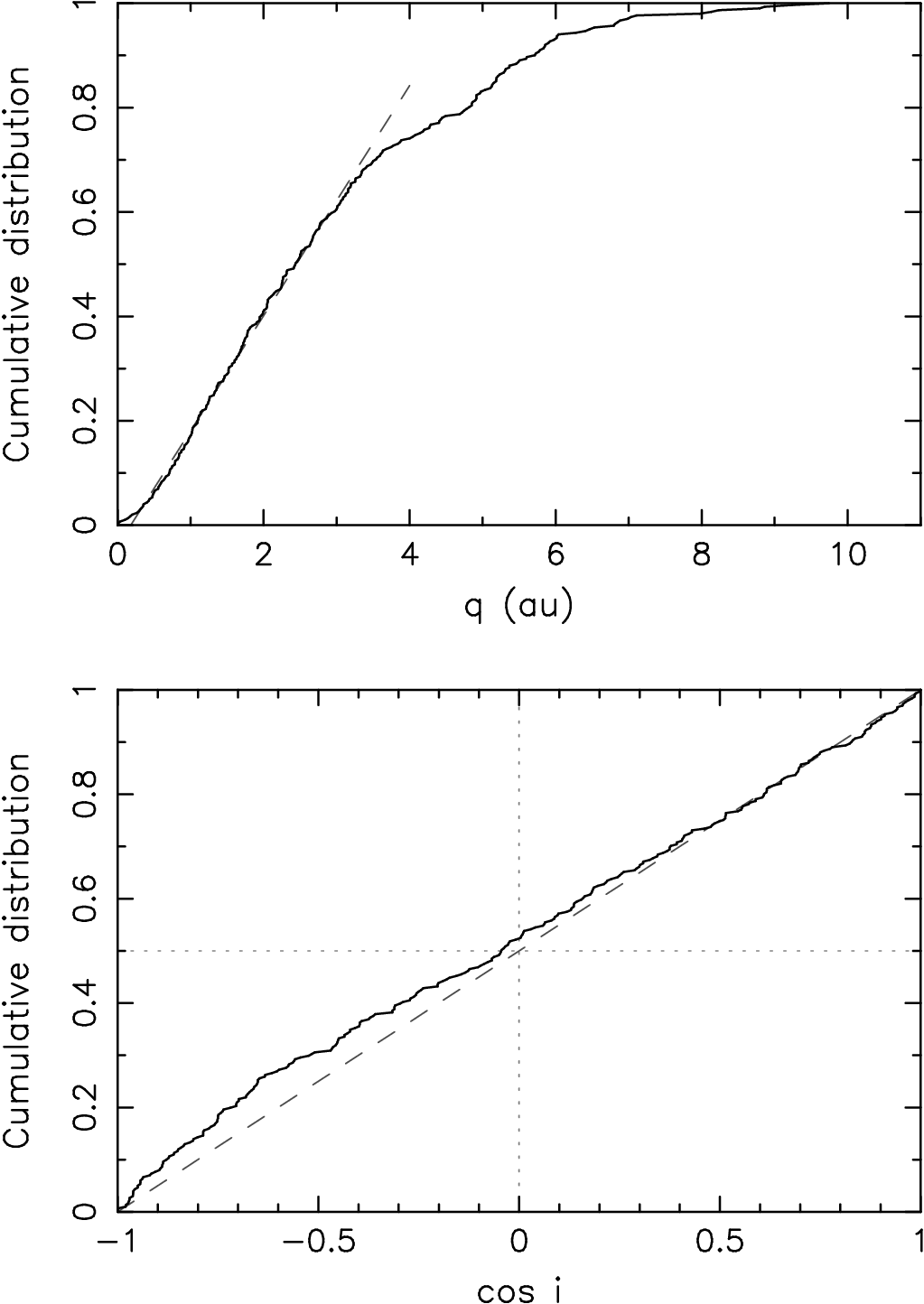}
\caption{Cumulative distribution of perihelion distance $q$ (top) and
 cosine of inclination $\cos i$ with respect to the ecliptic plane
 (bottom) for the selected sample of $318$ 1A and 1B orbits in
 MWC08. The gray dashed line in the upper panel shows a linear
 approximation for $q\leq 3$~au for reference. The dashed gray line
 in the bottom panel corresponds to an isotropic distribution (dotted
 lines indicate polar orbits, $\cos i = 0$, the median value for an
 isotropic distribution).}
 \label{mw_qi}
\end{figure}

Given the wealth of data in MWC08, and being cautious about the biases
mentioned above, we opted to analyze only the 1A- and 1B-flagged
orbits \citep[see, e.g.,] []{marsden1978}. This is a subset of 318
comets with the most accurately determined orbits in the
catalog. Figure~\ref{mw_a} shows the distribution of semimajor axis
$a$ of this sample of MWC08 comets. Here we use ${\rm log}\,a$ as the
abscissa instead of $1/a$, which is more suitable to study the
sub-class of comets on nearly parabolic orbits (Sec.~\ref{war}). This
choice allows us to distinguish the population of returning comets
with $a\lesssim 15000$~au from those from the canonical Oort peak with
$a\gtrsim 15000$~au. We shall also occasionally denote the latter
group as new comets, although both previous work
\citep[e.g.,][]{kq2009,DK2011,KD2013,DK2015,KD2017} and our
integrations show that a number of observed LPCs with $a\gtrsim
15000$~au have visited the planetary zone before. The fraction of
observed LPCs in the Oort spike is $\simeq 37$\% \citep[also see][who
used the 1993 edition of the Marsden-Williams catalog of
LPCs]{WT99}.

Figure~\ref{mw_qi} shows the cumulative distribution of perihelion
distance $q$ and cosine of inclination $\cos i$ for the sample of
$318$ new and returning comets from MWC08. The perihelion distribution
is fairly well-matched by a linear fit up to $q\simeq 3$~au, with
perhaps only a slight deficiency of the lowest-$q$ orbits ($\leq
0.2$~au, say). Beyond $3$~au, the distribution diverges
from the linear trend and becomes shallower, likely due to biases in
the data set (i.e., comets with larger perihelion distances are
typically fainter and thus harder to discover). However, if we were to
restrict ourselves to the subset of about 130 comets in the Oort peak
($a>12500$~au, say), the $q$- and $\cos i$-distributions would be
consistent with those given in Fig.~\ref{w_qi}. In particular, the
linear part of the $q$-distribution would extend to nearly $q\simeq
6$~au. We thus interpret the missing population of comets beyond
$q\simeq 3$~au in the upper panel of Fig.~\ref{mw_qi} primarily as a
deficiency of returning comets, perhaps due to fading of their
brightness in subsequent returns. On a physically deeper level, such a
fading pattern may result because the returning comets already
exhausted their content of supervolatiles, which might have driven
their huge activity on their first appearance. When these comets
return, it may be primarily the water sublimation below $\simeq 3$~au
which triggers their activity. Beyond Jupiter's orbit, even new comets
may be too faint to be detected by available surveys; only a small
fraction of the known population of LPCs has $q > 5$~au. There are
also biases subtler than the obvious lack of large-perihelion
comets. Note, for instance, that the linear progression of the
cumulative $q$-distribution is expected at the crudest approximation
\citep[e.g.,][pp.~127-130]{F2005}. Nevertheless, numerical models
that take planetary perturbations into account \citep[e.g.,][and
Sec.~\ref{res_1} below]{WT99,fetalAA2017} predict a slightly
nonlinear progression. This is not seen in the upper panel of
Fig.~\ref{mw_qi}, possibly because: (i) some comets are missing in the
MWC08 sample even below $q\simeq 3$~au, and/or (ii) the sample is not
homogenized to a common absolute brightness limit, such that a certain
number of smaller (and intrinsically less bright) comets contribute at
small $q$ values. We do not feel comfortable removing either of these
possible effects.
\begin{figure}[t!]
\epsscale{1.1}
\plotone{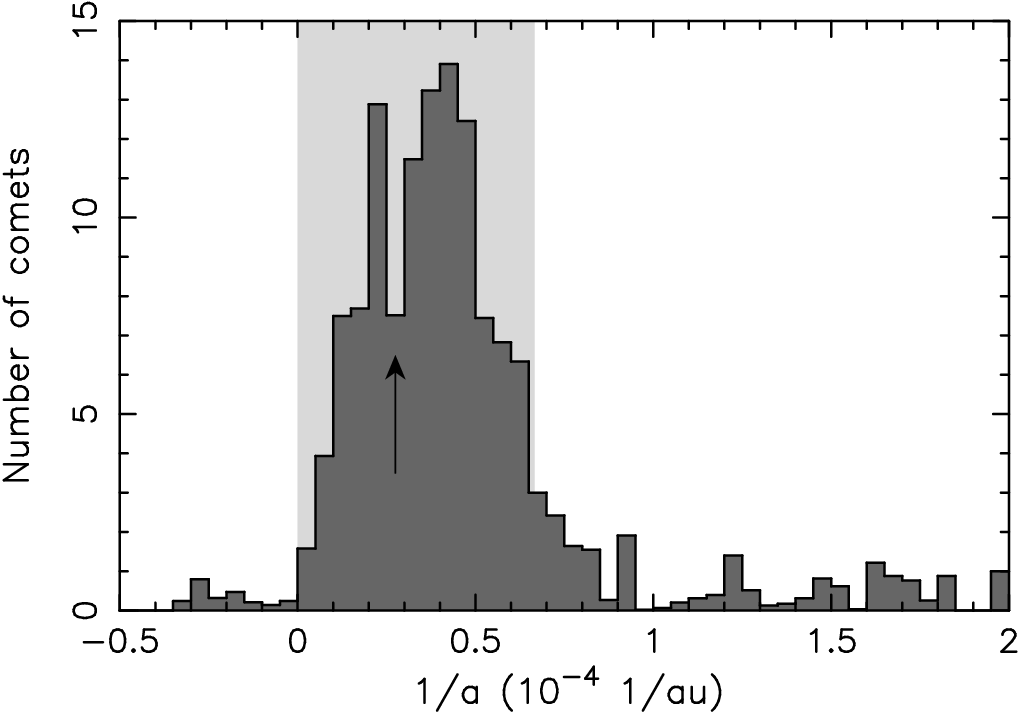}
\caption{Distribution of binding energy for LPCs on nearly parabolic
 orbits expressed as $1/a$ values (positive for elliptic orbits,
 negative for hyperbolic orbits). We use 134 entries in the
 Kr\'olikowska et~al.\ catalogs for which the stated uncertainty in
 $1/a$ is smaller that $10^{-5}$~au$^{-1}$ (this limiting value is
 twice as large as the bin size used). Each comet is represented by a
 Gaussian distribution with mean equal to the nominal value of $1/a$
 and standard deviation of the uncertainty in $1/a$. The gray
 rectangle highlights what is traditionally described as the Oort
 peak ($a\gtrsim 15000$~au here).}
 \label{w_a}
\end{figure}
\begin{figure}[t!]
\epsscale{1.1}
\plotone{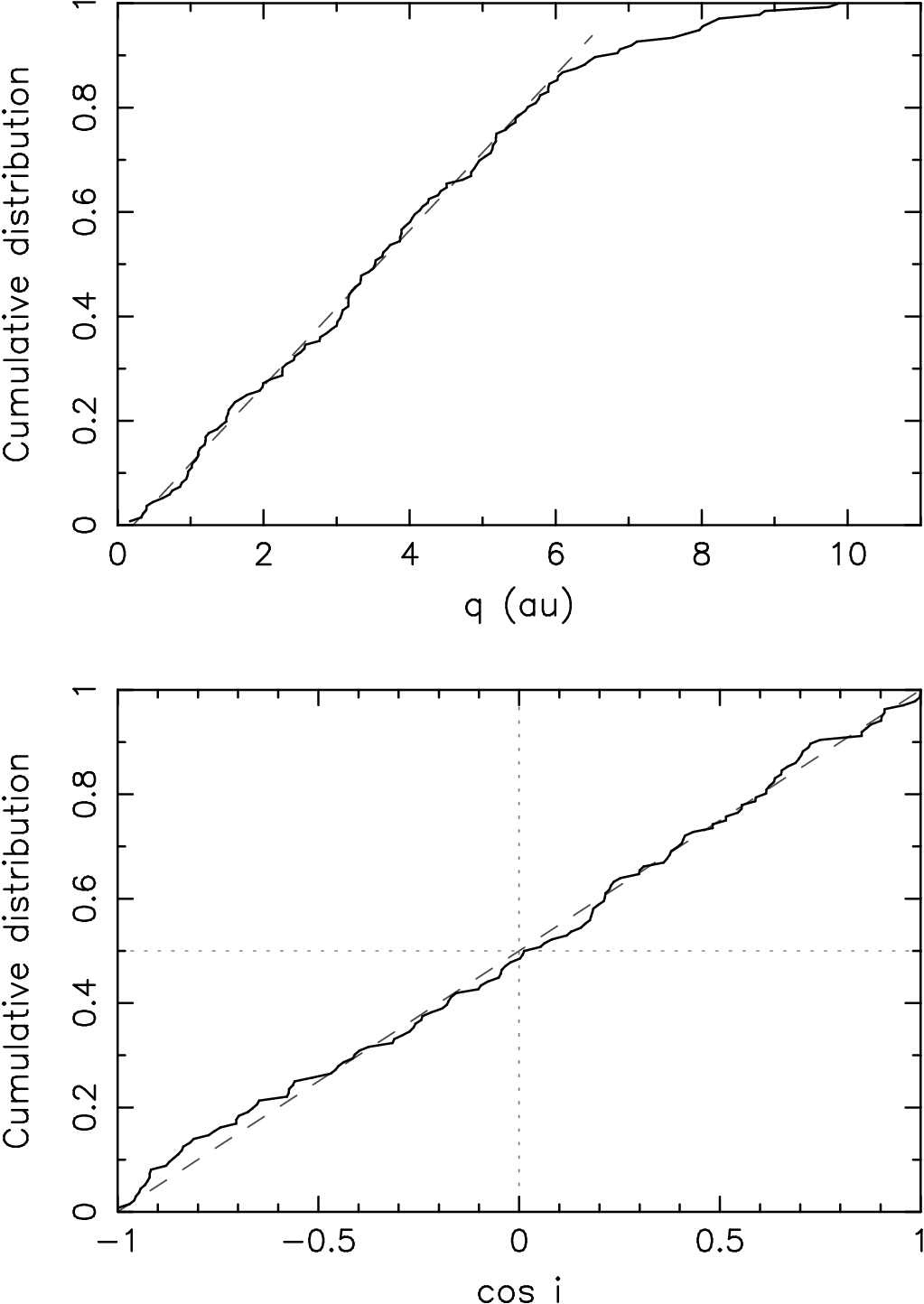}
\caption{Cumulative distribution of perihelion distance $q$ (top) and
 cosine of inclination $\cos i$ with respect to the ecliptic plane
 (bottom) for the selected sample of $134$ accurate orbits in the
 Kr\'olikowska et~al.\ catalogs of LPCs on nearly parabolic
 orbits. The dashed gray line in the upper panel shows a linear
 approximation for $q\leq 6$~au for reference. The dashed and gray
 line in the bottom panel correspond to an ideally isotropic
 distribution (dotted lines indicate polar orbits, $\cos i = 0$,
 the median value for an isotropic distribution).}
 \label{w_qi}
\end{figure}

The inclination distribution seen in the lower panel of
Fig.~\ref{mw_qi} is basically isotropic with only a slight excess of
retrograde cases. Again, when only the Oort peak comets of the LPCs in
MWC08 are used, the inclination distribution becomes closer to that of
an isotropic population. We thus believe that the small excess of
retrograde orbits originates primarily from the returning population
of LPCs.

\subsection{Orbital characteristics of nearly parabolic comets}\label{war}
As mentioned above, in order to describe comets on nearly parabolic
orbits in the Oort peak, we use data collected by a group of Polish
astronomers led by Kr\'olikowska. This represents a union of data
published in \citet{K2014}, \citet{ketal2014} and \citet{KD2017},
altogether 186 comets. Each entry in this catalog, as used here,
represents the orbital parameters of the original orbit together with
the estimated uncertainty.

Figure~\ref{w_a} shows the distribution of semimajor axis $a$ as a
function of $1/a$. Given a sufficiently large number of entries in the
catalog, we again restricted ourselves to a set of the most accurately
determined orbits. Here we only use those for which the uncertainty in
$1/a$ does not exceed $10^{-5}$~au$^{-1}$, thus reducing the sample to
134 comets. According to methods in \citet{K2014}, and the following
papers in their series, we represent each comet with a Gaussian having
the mean and standard deviation from the catalog. These data were then
represented as a histogram with bin size $5\times 10^{-6}$~au$^{-1}$,
about the median uncertainty of the cometary data. The data show the
structure of the Oort peak in a great deal of detail. \citet{KD2017}
note the division of the distribution by a dip at about $a\simeq
40000$~au (see the arrow in Fig.~\ref{w_a}), and associate it with a
separation of dynamically new and old orbits (see also Section~\ref{npc}).

Figure~\ref{w_qi} shows the cumulative distribution of perihelion
distance $q$ (top) and cosine of inclination $\cos i$ (bottom) for the
selected sample of 134 nearly parabolic comets from the Kr\'olikowska
et~al.\ catalogs. When compared with Fig.~\ref{mw_qi}, the behavior is
now simpler: (i) the linear trend in $q$ continues to nearly $6$~au,
before falling below the line, and (ii) the inclination distribution
closely matches an isotropic population, with only small
fluctuations. However, more subtle biases, such as the missing
expected nonlinear contribution in the $q$-distribution discussed in
the previous section, may still be present.

\subsection{Cometary flux and size distribution}\label{flu}
Unlike asteroids, comets hide the sizes of their nuclei with a huge
range of activity when they become observable. This brings large
difficulties in understanding their population parameters, in
particular their size distribution and/or size-limited flux.

Comets' intrinsic brightness is usually expressed in terms of the
absolute total magnitude $H$, which is related to the apparent
magnitude $m$ using a relation $H=m-5\,{\rm log}_{10}\,d-2.5\,n\,{\rm
log}_{10}\,r$ \citep[e.g.,][]{F2005}. (Cometary absolute magnitude
determinations sometimes include a term that accounts for non-zero
solar phase angle; we ignore this correction.) Here $d$ and $r$ are
the geocentric and heliocentric distances, respectively, and $n$ is
the photometric index, which strongly depends on the strength and
nature of a given comet's activity. For an inactive (asteroidal) body,
$n = 2$. Often $n=4$ is assumed for comets, leading to the
conventional absolute magnitude $H_{\rm 10}$. However, comets show a
great diversity in their activity and $n$ indexes ranging from $1$ to
$10$ have been reported for different comets \citep[with even more
extreme values on occasion, e.g.,][]{w1978}. Additionally, in many
cases photometric observations are not available for a large enough
interval of heliocentric distances $r$, so that the $n$ value of a
given comet is unknown. In this situation, $H_{\rm 10}$ is canonically
considered as the cometary absolute magnitude and taken as a proxy for
a physically more justified value of $H$. One should then understand
that such values may cause significant biases.

Yet another difficulty stems from the relation between the absolute
magnitude $H$ and the nucleus diameter $D$. This is because in nearly
all situations the observed brightness of a long-period comet 
results from sunlight reflected by its large coma with basically no,
or very little, contribution from the nucleus. Subtraction of the coma
is a tricky business \citep[see, e.g.,][]{Hui2018}. 

To circumvent these troubles, \citet{sf2011} used a determination of
non-gravitational forces in the motion of a sample of LPCs with
$q<2$~au to infer their nuclear masses. By assuming a mean bulk
density of $0.4$~g~cm$^{-3}$, they were able to estimate the effective
sizes of the nuclei. Running this analysis for a sample of $15$
well-observed LPCs, \citet{sf2011} were able to find an approximate
relation between $H$ and $D$ for this class of comets: $\log_{10} D
\simeq 1.2 - 0.13\,H$ [Note, however, that other authors have obtained 
similar relationships with different constants on the right hand side;
see the review in \citet{F2005}. If the light reflected by a comet 
is proportional to $D^n$, where $n$ is a constant, the coefficient of 
$H$ is $-0.4/n$. Thus the relation found by \cite{sf2011} implies 
$n\approx 3$, i.e., the reflected light is proportional to the volume 
of the nucleus, not its surface area. \cite{Weissman1990} finds, based 
on 1P/Halley, that $\log_{10} D \simeq 1.9 - 0.13\,H$ (for a density of
$0.4$~g~cm$^{-3}$), which implies that comets are $\approx 5$
times bigger than one obtains using the \cite{sf2011} relation.]
As an example, an $H=11$~magnitude comet would have, using the relation
of \citet{sf2011}, $D\simeq 600$~m. \citet{fs2012} used this analysis
to infer that the size distribution of active LPCs may be shallow for
$D \gtrsim 4.8$~km, steep between $\simeq 2.8$~km and $\simeq
4.8$~km, and shallow again between 1.2 and 2.8~km [for 1.2--2.8~km,
$N(>D) \propto D^{-1.54 \pm 0.15}$, 
where $N(D)$ is the cumulative number of nuclei with
diameter larger than $D$], and even shallower for smaller nuclei.  A
possible caveat, not accounted for in the uncertainty budget, is that
the analysis of \citet{sf2011} depends on the shape and location of
active areas on the cometary nucleus. These factors are highly
uncertain, especially for LPCs, and might affect their results.

Another, in principle more accurate, method would be to observe comets
at very large heliocentric distances in both visible and infrared
bands.  Assuming no, or very small, activity, one could run
traditional analysis known from asteroidal studies to determine
nuclear size. Alternatively, if observations are performed at smaller
heliocentric distances, one may hope to characterize the cometary
activity well enough to be able to subtract it from the total
fluxes. With that method the signal of the nucleus would be
obtained. Such an approach was conducted by \citet{betal2017}, who
used NEOWISE observations of a sample of 20 LPCs to infer their
sizes. They found a shallow [$N(>D) \propto D^{-1.0 \pm 0.1}$]
cumulative size distribution for LPCs between $\approx 1$ and 20~km in
diameter.


The differences mentioned above show that issues regarding the size
distribution of LPCs are still far from being resolved. In this
situation, we will not try to match details of the size distribution
of our studied sample of comets. Rather, we shall satisfy ourselves
with grossly matching the flux of LPCs above some size limit and below
some perihelion distance with our model. Based on observations by the
Lincoln Near-Earth Asteroid Survey (LINEAR), \citet{francis2005}
estimated an annual flux of about $11$ LPCs (dynamically new and old)
with $q<4$~au and absolute magnitude $H < 10.9$. (This range of 
absolute magnitudes corresponds to
cometary diameters $\gtrsim 1.0$~km and 2.4~km, respectively, for the
magnitude-mass relationships of \cite{Bailey1988} and \cite{Weissman1990} 
and nucleus density of $0.6$~g~cm$^{-3}$ that \citet{francis2005} uses.)
This result is sometimes also expressed as
a flux of 4 dynamically new comets with $q<5$~au and absolute
magnitude $H\lesssim 11$ per year \citep[e.g.,][where dynamically
new comets are roughly characterized with $a>10000$~au]{fetalAA2017}. 
This correspondence stems from (i) the
approximately linear cumulative distribution of LPCs with perihelion
distance $q$ (Secs.~2.1 and 2.2), and (ii) the assumption that
dynamically new comets represent about 1/3 of all LPCs \citep[Sec.~2.2
and][]{fs2012}.

To show that even the LPC flux estimate is not known accurately, we
note that the analysis of NEOWISE data by \citet{betal2017} obtained 
$\simeq 7$ LPCs larger than $1$~km passing annually within $1.5$~au 
from the Sun, which they stated to be about $2.6$ times larger than 
the result of \citet{francis2005}. This indicates that systematic
errors are still present in studies of LPCs. At present, obtaining a
rough correspondence (within a factor of a few) should be considered as
a satisfactory result. 


\section{Numerical model of LPCs}\label{model}
The initial orbital distribution for comets in our model is tightly
linked to the formation of the giant planets and their orbital
evolution in the early Solar system. The planets are assumed to emerge
from the gas-dominated infancy phase of the nebula in a compact, most
likely resonant, configuration, and further evolve orbitally due to
interactions with leftover planetesimals.  The solids which are
roaming on planet-crossing orbits are quickly removed, causing
(initially slow) orbital evolution of the planets. However, a huge
reservoir of planetesimals exterior to the orbit of Neptune remains
mostly intact for some time. The outer planetesimal disk, with an
estimated total mass of $\simeq 20$ Earth masses, is at first slowly
eroded at its inner edge, providing fuel for the planets' continuous,
slow migration. According to current knowledge, though, the
tightly-packed planet configuration became unstable and underwent
reconfiguration \citep[a modern version of this scenario is often
called the Nice model; e.g.,][]{nice}. As a consequence of this
chaotic and violent phase, Neptune entered the outer planetesimal
disk, proceeded to the outer edge of the dense part of the disk at
$\simeq 30$~au, and within $\simeq 100$ caused its entire dispersal. 
Most of the planetesimals were ejected from the Solar
system, some impacted the Sun and planets, and some ended up in
various long-lived reservoirs of small bodies in the Solar
system. With about $\simeq (4-6)$\% probability, the Oort cloud is by
far the largest surviving population of planetesimals \citep[see][and
Sec.~\ref{our_cloud} below]{dones2004,bm2013,nesvornySPC}. The
other end states have much smaller probabilities, such as: (i) $\simeq
1.5\times 10^{-4}$ for Plutinos in the exterior 3:2 mean motion
resonance with Neptune, $\simeq 5\times 10^{-4}$ for the hot
population of the classical Kuiper belt
\citep[e.g.,][]{nesvorny2015b,nv2016}, (ii) $\simeq 3\times 10^{-3}$
for scattering disk objects \citep[e.g.,][]{nvr2016,nesvornySPC},
(iii) $\simeq (5-8)\times 10^{-6}$ for the asteroid belt
\citep[e.g.,][]{levison2009,abelt}, (iv) $\simeq (2-3)\times 10^{-8}$
for irregular satellites around Jupiter, Uranus and Neptune and about
twice as large for those about Saturn \citep[e.g.,][]{irregs}, and (v)
$\simeq (5-7)\times 10^{-7}$ for Hilda and Trojan populations in the
3:2 and 1:1 mean motion resonances with Jupiter \citep[e.g.,][]{trojans,abelt}.

Unlike in the case of the Oort cloud, bodies in these other
populations of small bodies are directly observable. These successful
applications of the model represent justification of its consistency,
but -- most importantly -- they allow us to calibrate it in a
quantitative way. This is because the population of Jupiter Trojans,
in particular, is very well observationally characterized from the
size of its largest members of $\simeq 200$~km down to a size of
$\simeq 2-5$~km \citep[e.g.,][]{getal2011,wb2015,yt2017}. Because 
Trojans underwent little collisional evolution after their implantation, 
at least for the observed sizes \citep[e.g.,][]{roz2016}, their current
population, together with the known
implantation probability, allows us to quantitatively calibrate the
original planetesimal disk population. Other, slightly more uncertain,
quantitative constraints are summarized in \citet{nv2016}. For the
model to be self-consistent, we thus use the previously determined
quantitative calibration and apply it to other populations of small
bodies for which the implantation probabilities were determined.

Before we comment on several particular modeling details in the
following sections, we summarize the primary strengths of our
beginning-to-end approach:
\begin{itemize}
\item our starting initial orbits for comets are arguably consistent
  with their original birth configuration;
\item our model builds all structures of the Oort cloud as a response
  to the adopted planetary evolution scenario;
\item the population in the Oort cloud, acting as a source for LPCs,
  is independently calibrated by constraints from the original
  planetesimal disk.
\end{itemize}
Note that we successfully used this method to study Jupiter-family and
Halley-type comets in \citet{nesvornySPC}. Here we apply it to the
case of LPCs. All that said, we admit that our model is far from being
perfect. Some of its main caveats are summarized in Sec.~\ref{cav}.

\subsection{Integration method}
While the work of \citet{nice} represents now an archetype, inaccurate
in several aspects, the Nice family of scenarios for early planet
migration has undergone further development in the past decade. Here
we use the class of five-planet models presented and tested in
\citet{nm2012} \citep[also see][]{bat2012}. It would have been ideal
to repeat some of their successful simulations with myriads of disk
particles, but this approach is not possible computationally.
Instead, we adopt the approximation of planet migration introduced in
\citet{nesvorny2015a,nesvorny2015b} and \citet{nv2016}. It is
important to point out that our runs here, except for issues of
exporting information about particle orbits and slightly different
stellar encounter files, are essentially identical with those in
\citet{nesvornySPC}. This makes a common basis for modeling orbits of
{\it all} comets, both short- and long-period, in our approach.

Jupiter and Saturn are placed on their current orbits (assumed fixed
at all times; terrestrial planets are not included in our
simulations). Uranus and Neptune start initially on orbits interior to
their current values and both are migrated outwards. In particular,
Uranus's and Neptune's initial orbits were circular with semimajor
axes $17$~au and $24$~au, both located in the Laplace plane defined by
Jupiter and Saturn. We use the {\tt swift\_rmvs4} code, part of the
Swift N-body package \citep[e.g.,][]{ld1994}, in which fictitious
forces were introduced to mimic radial migration, eccentricity and
inclination damping of the orbits of Uranus and Neptune. These forces
are parametrized by exponential timescales, as discussed in
\citet{nv2016}. For instance, Neptune's semimajor axis asymptotically
approaches its current value of $30.11$~au, while its eccentricity and
inclination are driven to zero. Similarly, Uranus is forced to
approach its current orbit. We assume a characteristic timescale
$\tau$ for these dynamical effects, common to all three elements (we
found no need to distinguish the effects on semimajor axis,
eccentricity, and inclination). Motivated by the full-fledged
simulations in \citet{nm2012}, we distinguish two phases of planetary
migration, separated by an instability when Neptune's orbit reaches a
heliocentric distance of roughly $27.8$~au. At that moment, Neptune's
orbit is assumed to undergo a slight discontinuity in its semimajor
axis due to encounters with the fifth giant planet \citep[this helps
to explain existence of the kernel in the Kuiper belt;
see][]{nesvorny2015a}. \citet{nm2012} also found that the migration
timescales differ slightly before and after the instability, typically
being shorter before and longer after. As discussed in \citet{nv2016},
$\tau_1=10$~Myr and $\tau_1=30$~Myr roughly bracket the range before
the instability, while $\tau_2=30$~Myr and $\tau_2=100$~Myr represent
the range after the instability (lower values correlate with an
initially more massive planetesimal disk and vice versa). The longer
timescales, especially $\tau_2$ after the instability, provide
somewhat better results. For example, they help to explain the
inclination distribution of the hot population in the Kuiper belt
\citep[e.g.,][]{nesvorny2015b,nv2016} and facilitate capture of
Saturn's spin axis into the $s_8$ secular resonance
\citep[e.g.,][]{vn2015}. While these details may not be crucial for
our study here, we run two sets of simulations: (i) case 1 ({\tt C1})
with $\tau_1=30$~Myr and $\tau_2=100$~Myr, and (ii) case 2 ({\tt C2})
with $\tau_1=10$~Myr and $\tau_2=30$~Myr. This is the same approach
chosen in \citet{nesvornySPC}.

The initial phase of planetary evolution, with their migration
implemented as above, is carried to $500$~Myr from the beginning. Both
Uranus and Neptune are at that moment very close to their current
orbits. From then on, we continue the integration without the
fictitious accelerations, taking into account only mutual
gravitational effects between the Sun and planets. This second phase
continues for $4$~Gyr. Therefore, at the end of our simulation its
timescale reaches $4.5$~Gyr, the approximate age of the Solar system.
This is important for correctly reproducing the extent, and comet
density, of all structures of the Oort cloud.

All integrations were performed with a time step of $0.5$~yr, but, as
explained in \citet{nesvornySPC}, we compared with limited runs using
shorter time steps to make sure the results were satisfactory. Only in
the last Gyr, between $3.5$~Gyr and $4.5$~Gyr, did we use a shorter
time step of $0.2$~yr. This is because we wanted to make sure the
integration allowed us to precisely determine the cometary state near
perihelion passage, as explained in Sec.~\ref{out}.

\subsection{Initial data: planetesimal disk}
Aside from the planets, our simulations propagate the orbits of a
large number of planetesimals in the initially trans-Neptunian
disk. These particles are assumed massless. In spite of their
collective mass of $\simeq (15-20)$ Earth masses, we thus neglect
their direct effect on the motion of the planets.  Nevertheless, since
the orbits of the planets are made to behave as in the more complete
simulations in \citet{nm2012}, which do include this feedback, this is
not a problem. We also neglect the self-gravity effects of the disk
particles with each other.

The planetesimal disk is assumed to have two parts: (i) a high-mass
part, initially extending from the orbit of Neptune to a heliocentric
distance of $\simeq 30$~au, and (ii) a low-mass extension to a
heliocentric distance of $\simeq 45$~au. In this work, as in
\citet{nesvornySPC}, we include only the massive part (i). This is
because only bodies from this part of the disk have a chance of
undergoing close encounters with the migrating Neptune and the other
giant planets, and thus to be efficiently transferred to various
small-body populations in the outer Solar system, such as the
scattered disk and the Oort cloud \citep[also
see][]{dones2004,detal2015}. Planetesimals from the outer part of
the disk, beyond $30$~au, may also contribute via subtle dynamical
effects (such as resonances), but the probability is low and the outer
disk has a small mass. Both indicate that the importance of the outer
disk is minimal.

Each of our simulations initially included one million disk particles
distributed from Neptune's orbit to a heliocentric distance of
$30$~au. The disk is assumed axisymmetric with a radial surface
density $\propto 1/r$.  Initial eccentricities and inclinations of the
disk particles are assumed to be very small, satisfying Rayleigh
distributions with standard deviations of $0.05$ and $2^\circ$,
respectively. Planetesimals are propagated in our simulations until
the final epoch of $4.5$~Gyr unless one of several elimination
conditions is satisfied: impact with the Sun or a planet, impact with
a passing star, or ejection from the Solar system. The latter is
assumed to happen when the heliocentric distance of the particle
exceeds $500000$~au.

\subsection{Galactic tide model}
Modeling the source regions of long-period comets, located in the
outskirts of the Solar system, requires including gravitational
effects from the Galaxy. These have two components: (i) the collective
effect of the global mass distribution in the Galaxy, resulting in a
smooth potential, and (ii) the impulsive, short-range effect of stars
passing very close to, or even through, the Oort cloud. We start with
the former, leaving description of the latter to the next Section.

We consider the simplest model of the galactic potential (see further
comments in Sec.~\ref{cav}). The Sun is assumed to move about the
center of the Galaxy on a constant circular orbit located in the
galactic midplane. The galactic potential is approximated with an
axisymmetric model, and in the solar neighborhood we approximate it as
a quadrupole. With this crude approach we can describe the associated
acceleration ${\bf f}$ in the motion of all bodies in our simulations
as follows. Assume a Sun-centered, slowly rotating orthonormal
reference frame $({\bf e}_x,{\bf e}_y, {\bf e}_z)$, such that ${\bf
  e}_x$ is oriented in a radial direction away from the center of the
Galaxy, ${\bf e}_y$ is transverse along the direction of solar motion
in the Galaxy, and ${\bf e}_z$ is normal to the galactic midplane. In
the quadrupole approximation ${\bf f}$ is a linear function of the
coordinates $(x,y,z)$. Traditionally, these are expressed in the form
\citep[e.g.,][]{ht1986,bt2008}
\begin{equation}
 {\bf f} = \Omega_0^2\left[\left(1-2\delta\right)x\,{\bf e}_x-y\,{\bf
  e}_y -\left(\frac{4\pi G\rho_0}{\Omega_0^2}-2\delta\right)z\,{\bf
  e}_z \right]\, , \label{tide}
\end{equation}
where $\delta = -(A+B)/(A-B)\simeq -0.09$, $\Omega_0 = A-B \simeq
2.78\times 10^{-8}$~yr$^{-1}$ and $\rho_0\simeq 0.15$~M$_\odot$
pc$^{-3}$. Here we adopted $A\simeq 14.82$~km~s$^{-1}$~kpc$^{-1}$ and
$B\simeq -12.37$~km~s$^{-1}$~kpc$^{-1}$ based on Hipparcos satellite
measurements of galactic Cepheids \citep{fw1997}; $A$ and $B$ are
the Oort constants and $\rho_0$ is the mass density in the solar
neighborhood. Recent re-evaluations of local galactic dynamics may
indicate a slightly larger $\delta$ value \citep[and small deviations
from axisymmetry, e.g.,][]{bovy2017}, but this is of minor
importance. The right hand side in Eq.~(\ref{tide}) is dominated by an
order of magnitude by the third term, which is proportional to
$\rho_0$. Visible matter contributes $\simeq 0.10$~M$_\odot$ pc$^{-3}$
\citep[e.g.,][]{bt2008,wdb2010}. The contribution of dark matter is
quite uncertain \citep[e.g.,][]{wdb2010,bt2012}. Our assumed increase
to $0.15$~M$_\odot$ pc$^{-3}$ is rather conservative and may even
overestimate the effective, long-term value of $\rho_0$. This may have
interesting implications, as we discuss in Sec.~\ref{concl}.

Stationarity and axisymmetry of the local galactic potential are
certainly large simplifications. Even if both applied to the total
potential of the Galaxy, the stationarity may be broken locally by the
Sun's oscillations about its roughly circular orbit. For instance, the
shorter of the radial ($x$) and vertical ($z$) periods is 
that of the vertical oscillations
$\sqrt{\pi/G\rho_0}$. The effective density of matter felt by the
solar neighborhood should oscillate with half of this period, some
$30$~Myr. Since the Sun is currently very close to the galactic
midplane, where density is maximum, the long-term average $\rho_0$ may
again be slightly smaller than assumed in our simulations. Detailed
analysis of such effects is, however, beyond the scope of this paper
\citep[see, e.g.,][]{getal2011}.

Our simulations use an inertial reference system with the $(x,y)$
plane defined by the invariant plane of the Solar system. Therefore,
we need to apply an appropriate transformation of ${\bf f}$ in
(\ref{tide}). This is simply achieved in two steps: (i) a slow
rotation about the $z$ direction with frequency $\Omega_0$, and (ii) a
fixed $\simeq 62.5^\circ$ tilt between the galactic and invariant
planes.

\subsection{Perturbations from stellar encounters}
Since the work of \citet{oort1950}, the role of perturbations from
individual stellar encounters has been discussed in the context of
cometary origin, in particular for LPCs. While opinion on the
prevailing driver (tides or stellar encounters) to bring comets into
the observable zone has varied, the present view highlights a
synergistic effect of both \citep[see,
  e.g.,][]{retal2008,fetalAA2011,fetalIca2011}. We thus include the
effects of stellar fly-bys in our simulations, though -- as in the
case of the tides -- we make important simplifications.

Results from the Gaia project will determine, no doubt, the state of
the art in defining the rate at which different stellar types/classes
presently encounter the Solar system. Data from the first and second
releases have begun to flow
\citep[e.g.,][]{bd2016,bj2018,bjetal2018}. However, up to this moment
no comprehensive compilation and debiasing of the data has been
published.  For that reason, our primary source is the work of
\citet{gsetal2001}, who analyzed data from the Hipparcos
mission. While more limited than the Gaia data, we believe that the
Hipparcos data are adequate for our purposes.
\begin{deluxetable}{cccc}[t!]
\tablecaption{Parameters of our four simulations
 \label{tab1}}
\tablehead{\colhead{Designation} & \colhead{$\tau_1$} & \colhead{$\tau_2$} &
           \colhead{Reference} \\ [-1.5ex]
           \colhead{} & \colhead{(Myr)} & \colhead{(Myr)} & \colhead{}}
\startdata
 {\tt C1V1} & 30 & 100 & 1 \\
 {\tt C1V2} & 30 & 100 & 2 \\
 {\tt C2V1} & 10 &  30 & 1 \\
 {\tt C2V2} & 10 &  30 & 2 
\enddata
\tablecomments{The second and third columns give the assumed parameters of
 planetary migration timescale before ($\tau_1$) and after ($\tau_2$) the
 instability. The last column provides a reference for the stellar encounter
 model: 1 stands for \citet{retal2008}, 2 stands for \citet{mbetal2017}.}
\end{deluxetable}

We implemented the scheme developed and described in detail in Sec.~2
of \citet{retal2008}. Choosing an interval of time, $4.5$~Gyr in
our case, their method allows us to create a sequence of stellar
encounters with the Solar system whose statistical properties match
those determined in the work of \citet{gsetal2001}. In particular, for
thirteen stellar categories of a given specific mass, from low-mass
$0.21$~M$_\odot$ M-dwarfs to high-mass $9$~M$_\odot$ B-giants, one
obtains: (i) the flux into a region of 1~parsec (206265~au) distance
from the Sun, (ii) the mean stellar velocity with respect to the local
standard of rest, and (iii) the parameters of the velocity dispersion
with respect to the local standard of rest. With this information, we
create a random sequence of initial conditions of stellar entries into
the $1$~pc heliocentric zone. Each data point specifies (i) where and
when the star enters, (ii) its heliocentric velocity, and (iii) its
mass. Since the relative motion of the Sun and the star is very nearly
hyperbolic, we may also determine the closest approach to the Solar
system. The model based on the original recipe of \citet{retal2008} is
denoted {\tt V1}. In order to ensure that fixed masses of the objects
in stellar classes do not create artifacts, we also developed a second
model {\tt V2}, where, for each of the thirteen stellar categories, we
use a range of masses with a given power-law distribution. These data
are taken from \citet{mbetal2017}.

Ideally, we would run a large number of simulations, where in the {\tt
V1} and {\tt V2} series of models, a random, and each time
different, sequence of stellar encounters would be taken into
account. However, each of our runs begins with one million particles
and is quite demanding of CPU time. As a result, we only performed one
of the {\tt V1} and {\tt V2} variants and combined them with cases~1
and 2 for planet migration described in Sec.~3.1. The complete set of
simulations is listed in Table~\ref{tab1}. While less than we would
wish, we note that we do not see any significant differences in the
results of our jobs (see Sec.~4). This in part justifies our limited
number of simulations.
\begin{figure*}[t!]
\epsscale{1.1}
\plotone{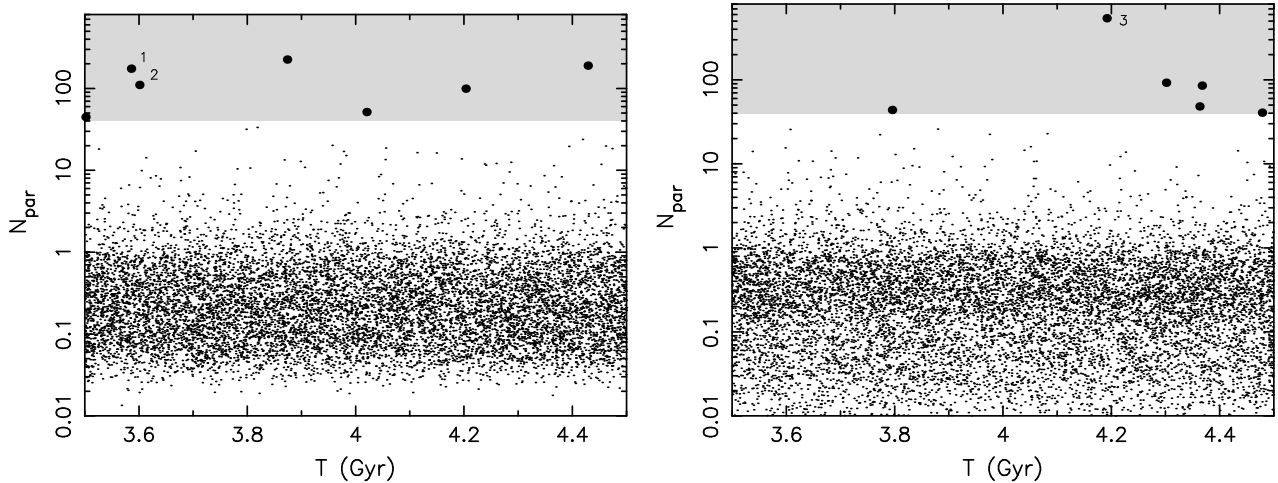}
\caption{An estimate of the number of comets $N_{\rm par}$ injected into
 the tidally active zone as a consequence of a stellar encounter
 \citep[see Eqs.~(1) and (2) in][]{fetalIca2017}: (i) left panel for
 {\tt V1} simulations, (ii) right panel for {\tt V2} simulations.
 Only the values in the last Gyr of the simulations are shown. The
 strong encounters with $N_{\rm par}\gtrsim 40$ in the gray box are
 highlighted by large symbols. Details on the comet showers associated with the
 encounters labeled 1, 2 and 3 are shown in Fig.~\ref{shower}. Most of
 the stellar encounters constitute a background with $N_{\rm par}\lesssim
 2-3$.}
 \label{nstar}
\end{figure*}

For sake of illustration, we find it useful to fold the
multi-dimensional information on the stellar encounters, such as their
mass, encounter velocity, and the closest approach, into a
single-parameter proxy. To that end we use $N_{\rm par}$ defined in
\citet{fetalIca2017} (their Eqs.~(1) and (2)).  According to this
source, $N_{\rm par}$ approximates the number of comets injected into
the observable region, and thus shows the importance of a given
encounter. We note that $N_{\rm par}$ is similar to a simpler $g$
parameter used in \citet{fbj2015}. The difference between the two
parameters occurs primarily for high-velocity encounters with low-mass
stars. However, since we use $N_{\rm par}$ only as an auxiliary
parameter to identify particularly important encounters, these
differences are not important. The real importance of the encounter is
further studied in Sec.~\ref{results} by tracing truly detectable
comets in our model.

Figure~\ref{nstar} shows $N_{\rm par}$ values in our single realizations of
the {\tt V1} and {\tt V2} encounter series in the last Gyr of the simulation.
Most of the values are $\lesssim 2-3$ and those constitute a background
signal. Occasionally, a star passes close enough to surpass this background.
The values are slightly more spread in the {\tt V2} model because of the
considered range of stellar masses. The highest values of $N_{\rm par}$ range
between $100$ and $\simeq 550$ in our simulations. Most often, these
correspond to subsolar-mass stars passing very close to the Solar system
and having small encounter velocities. Only one of these cases, labeled 2
on the left panel of Fig.~\ref{nstar}, corresponds to the encounter of a
$9$~M$_\odot$ giant star. We found that the encounters with $N_{\rm par}
\gtrsim 40$ (red symbols) produce observable comet showers in our
simulations (Sec.~\ref{results}).

The combined frequency, over all stellar types, of encounters within
$1$~pc of the Sun is $\simeq 11$ per Myr. This value seems realistic, even
slightly smaller than preliminarily inferred from the Gaia data
\citep[$19.7 \pm 2.2$ per Myr,][]{bjetal2018}. Obviously, this flux is
dominated by encounters with the lowest-mass dwarfs. The closest
generated approaches to the Sun over the $4.5$~Gyr time span were
$\simeq 1700$~au. These anomalous encounters penetrate not only the
outer, but also the inner, parts of the Oort cloud. However, because
the cumulative number of stellar encounters with perihelion smaller
than $q_\star$ scales as $\propto q_\star^2$, most of the encounters
are much more distant. For instance, their number with $q_\star
\lesssim 40000$~au is only $\simeq 4$\% of the total. It is also
interesting to note that these statistics fit the parameters of the
closest known stellar approach within the $\pm 10$~Myr interval of
time from the present: the dwarf star Gliese 710 is predicted to
approach within 10000--20000~au of the Sun about $1.3$~Myr from now
\citep[90\% confidence interval for distance,
e.g.,][]{bd2016,bj2018,bjetal2018}.

Having prepared a look-up table of the initial conditions of stars at
the $1$~pc sphere about the Solar system, the effect of stellar
encounters was incorporated in our simulations by adding the stars as
new massive bodies into the integrations. Because some of the stars
may spend up to a few hundred thousand years within $1$~pc of the Sun,
at moments the simulation may account for several passing stars. The
stars were followed throughout their encounters until they again
reached a distance of $1$~pc from the Sun.

\subsection{Comet production runs}\label{out}
The observational information about LPCs, summarized in Sec.~2, is
based on data collected over the past two centuries (although more
than 80\% of them are even more recent, and represent discoveries over
the past two to three decades only). Ideally, one would wish to
compare this data set to modeled comets during a comparably short
interval of time. For this to work, however, one would need to include
many more planetesimals in our simulations ($\simeq 10^{12}- 10^{13}$
instead of $10^6$; Sec.~\ref{results}). This is obviously impossible
for computational reasons.

In this situation, we need to trade the smaller number of integrated
planetesimals for a longer interval of time over which data are
collected \citep[also see][where a similar approach was
used]{nesvornySPC}. To compensate for the ``missing'' six to seven
orders of magnitude, we need a time interval of at least several tens
of Myr. In fact, to have enough statistics, we used the last Gyr in
our simulations for this purpose. We find this method adequate,
because the comet flux within this interval of time is approximately
steady (see, e.g., Figs.~\ref{nc_20} and \ref{nc_5}). In particular,
the very slow decline due to late erosion of the Oort cloud represents
an effect of only $\simeq 10-15$\% \citep[most of the population
dynamics of the Oort cloud is completed by $0.5-1$~Gyr after the
beginning; e.g.,][]{dones2004}. As a result, the underlying
assumption of a steady state of our model is only very weakly
violated. Following methods in \citet{fetalIca2017} and
\citet{fetalAA2017}, we only discard periods adjacent to the strongest
comet showers (roughly indicated by encounters having $N_{\rm par}\geq
40$; Fig.~\ref{nstar}). Collectively, these cover only about $\simeq
20-30$~Myr from the target Gyr interval of time. This is because our
analysis of the comet showers in Sec.~\ref{results} indicates that
their signal fades away within $2-5$~Myr (the analysis in
\citet{bj2018} or \citet{bjetal2018} showed no stellar encounter within
the past $\simeq 10$~Myr that could produce a noticeable comet shower).

Following these considerations, we modified the output from our
numerical code between the epochs $3.5$~Gyr and $4.5$~Gyr since the
beginning (the last Gyr). We monitored the heliocentric distances $r$
of all particles in our simulations. When, for a given particle, $r$
became smaller than 20~au, we followed its evolution and output the
planetary and particle state vectors near its perihelion. Because we
focus on LPCs in this paper, the output was performed only when the
heliocentric semimajor axis satisfied $a\geq 35$~au (i.e., orbital
period longer than $\approx 200$~yr).  After completing the primary
simulation, we used these data specific to LPCs in the last Gyr to
compute the particles' original orbits before they entered the
planetary region. In particular, we performed a sequence of short
integrations backward in time from each of the data files and followed
the orbit until it reached a heliocentric distance of $250$~au. If the
motion reached aphelion before this limit, we used the aphelion state
vectors. The dynamical state of the comet was then transformed into
the Solar system barycentric frame and the barycentric orbital
elements were computed. These are to be compared with the data
outlined in Sec.~2. Note that we aligned the 250~au limit with the
practice used in the Kr\'olikowska et~al.\ catalogs
\citep[e.g.,][]{ketal2014,K2014,KD2017}. The original orbits in MWC08
were computed at a smaller heliocentric distance, but the difference
is insignificant.

With the parameters described above, we have full control of the LPC
orbital evolution when their perihelia decrease below $20$~au. The
choice of this limit resulted from a compromise between several
factors. First, it allows us to learn about the orbital evolution even
before the comet becomes observable with current surveys (heliocentric
distances $\lesssim 10$~au; Sec.~2). Second, it also allows us to
theoretically characterize a putative population of LPCs with
perihelia between the orbits of Saturn and Uranus. This is
interesting because this population may 
be in reach of forthcoming surveys (note that today's catalogs contain
only four well-observed long-period comets with perihelia 
beyond Saturn's semi-major axis of $9.54$~au, with
C/2003 A2 (Gleason) ($q\simeq 11.43$~au) being the record holder).
Both reasons may motivate us to push the
limit even further than $20$~au, but this is problematic at the moment.
The population of LPCs steeply increases beyond the perihelion limit
of $\simeq 15$~au (see Figs.~\ref{nc_qdist} and \ref{nc_qa}). Therefore,
extending the target zone where comets are being monitored towards the
orbit of Neptune (i.e., $30$~au) would (i) produce increasing demands on
disk storage, and (ii) slow down the simulations.

\subsection{Fading problem for LPCs}\label{fad}
\citet{oort1950} noted that the observed energy distribution of LPCs,
which is sharply peaked for $1/a\leq 10^{-4}$~au$^{-1}$ (e.g.,
Fig.~\ref{w_a}), is only compatible with model predictions if comets
are allowed to remain observable only for a certain number of returns
to the inner Solar system. In particular, Oort postulated an average
disruption probability of $1.4$\% per perihelion passage. But even
with this assumption, he was unable to explain the sharp concentration
of comets on nearly parabolic orbits. Therefore he assumed that most
LPCs (some 80\%) are overly active when first arriving in the
observable region with small perihelion distances and, therefore,
exhaust most of their volatiles that feed the observable comae. When
the comets arrive again, they are much fainter and supposedly escape
detection. Whether they actually do arrive again, or disrupt
\citep[e.g.,][]{letal2002}, is not really relevant to our work.  Both
constitute what is called the comets' fading.

Comets are followed in our simulations as unbreakable point particles
and may suffer elimination only for dynamical reasons. Because it
would be inconvenient to implement the physical lifetime (fading)
effects in the numerical simulation of the orbital evolution, we save
it for post-processing of the results. This is possible because we
have information about the returns of the given comet before it was
dynamically eliminated (obviously, only within the target heliocentric
region of $20$~au).  Cometary fading may be approached as a physical
process with all its complexity. This is, however, quite beyond the
scope of this paper.  Rather, we shall adopt a simple, empirical
description of the fading process primarily as a function of the
number of returns to the Solar system.  A very nice overview of
possible choices is given in Sec.~5.5 of \citet{WT99}.

We first tried the simplest possible choice, namely to allow a certain
number of perihelion ($q$) returns below a $q$ limit $N_{\rm p}(q)$:
this option was used by \citet{nesvornySPC} for short-period comets
with a choice $q=2.5$~au.  However, we found that this provides
unsuitable results for LPCs, even when changing the $q$ limit. In
particular, the ratio of the number of new comets in the Oort peak to
the number of returning comets was never well satisfied \citep[this is
  in agreement with results in][]{WT99}. This failure is because such
a simple fading law does not fit Oort's original suggestion that LPCs
fade more at their first appearance, and much less later on. One could
try ad hoc assumptions about different fading probabilities at
different returns. At this point it is actually easier to assume some
simple smooth function of the return number. This parametrization was
introduced by \citet{whipple1962} and successfully used by
\citet{WT99}. For those reasons, we shall adopt the same approach
here.
\begin{figure}[t!]
\epsscale{1.1}
\plotone{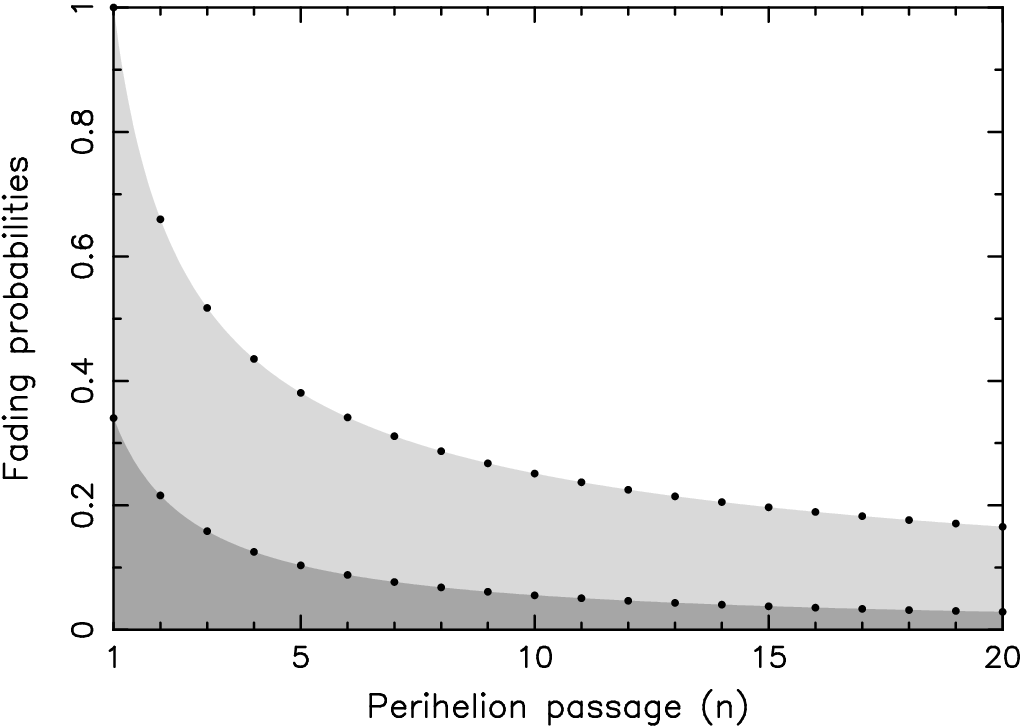}
\caption{An example of a one-parameter fading law used in our
 analysis: a power-law probability $\Phi_n=n^{-\kappa}$ that a
 dynamically new comet survives fading for at least $n$ perihelion
 passages \citep[$n=1$ means its first appearance;
 see][]{whipple1962,WT99}. The upper curve, enclosing the light
 gray area, shows $\Phi_n$ for $\kappa=0.6$ ($\Phi_1=1$ implies the
 comet has been observed). The bottom curve, enclosing the dark gray
 area, shows the conditional probability $\psi_n = 1 -(\frac{n}{n+1})^k$ that a comet which
 survived $n$ perihelion passages will fade before returning for the
 $(n+1)$th time.}
 \label{fading}
\end{figure}

\citet{whipple1962} assumed the probability $\Phi_n$ for a comet to
survive at least $n$ perihelion returns is a simple power-law
function: $\Phi_n=n^{-\kappa}$, where $\kappa$ is a constant.  Both
\citet{whipple1962} and \citet{WT99} found that $\kappa\simeq 0.6$
provides a good ratio between new and returning
LPCs. Figure~\ref{fading} shows the properties of this choice.
We note that some 35\% of comets survive only one return and
some 60\% of comets survive only 5 returns. Beyond that, however, the
survivability significantly improves, nearly as if there were two
categories of objects: some which die very quickly and some which have
very good chances of survival even after many returns. This is the
reason for the success of the empirical fading law suggested by
Whipple. At the same time, one should admit the limitations of this
single-parameter law. Its applicability up to now perhaps means that
the comets (i) are observed in a still rather limited region of
perihelion distances (note that \citet{WT99} limited their study to
$q<2.5$~au), and (ii) are mostly of a typical size $D$. In
principle, the fading must depend on both $q$ and $D$, such that
larger comets, and those passing at larger perihelia, should live
significantly longer. Some aspects of the size dependence in cometary
fading have been quantitatively documented, for instance, for
long-period comets with $q<0.5$~au by \citet{bortle1991} and LPCs with
$q<1$~au by \citet{Sekanina2019}
\citep[also see discussion in][]{whipple1992}. More, and especially 
well understood, observations will be needed to test the
complex parameter dependence of cometary fading.
In this paper, we stick with the original simple formulation of
\citet{whipple1962}.

\subsection{Features not included in our model}\label{cav}
Even though we made efforts to present a complete and consistent model
for the origin and evolution of LPCs, we neglect several important
elements. Here we briefly recall these caveats which will need to be
considered in future work.
\smallskip

\noindent{\it Effects of the solar birth cluster.-- }In all
likelihood, the Solar system was initially formed within an embedded 
cluster of stars \citep[see the reviews by][]{adams2010,
Pfalzner2015}. Various constraints imply that this birth environment
contained hundreds to perhaps a few thousand stars, all located within
a few parsec zone. Depending on the cluster parameters, a typical
solar analog could have left its natal cluster in a couple of tens of
Myr. Before reaching a more friendly environment characterized by the
current galactic tidal field and the current frequency of stellar
encounters (both outlined above), the early Solar system thus
experienced much more fierce conditions.

In terms of small body deposition in the trans-Neptunian zone, most
studies focused on two aspects: (i) formation of a fossilized inner
Oort cloud, possibly extending inward to a few hundred au from the
Sun, and (ii) implantation vs. erosion of the classical Oort
cloud. The first line of investigation was motivated by the discovery
of a population of extremely detached trans-Neptunian objects, such as
Sedna ($q\simeq 76$~au). Indeed, various simulations \citep[see,
e.g.,][]{fb2000,betal2006,kq2008,betal2012} have shown that stellar
encounters at very small distances, typical in the initial phases of
the cluster evolution, allow the Oort cloud to extend inward enough to
comfortably explain the existence of Sedna and similar bodies. This
structure would be unaffected by currently acting galactic tides and
thus would remain a fossil relic of the natal stage of the Solar
system. It would not contribute significantly to the currently 
observable population
of LPCs.  It may become a relevant source of a population of LPCs with
more distant perihelia, beyond the orbit of Saturn, if observed in the
future. However, some studies suggest that the fossilized inner
extension of the Oort cloud may actually be depleted in small bodies
(diameters less than $\simeq 4$~km). This is because gas drag in the
primordial solar nebula might have prevented transport of such small
bodies to this source zone \citep[e.g.,][]{betal2007}.

As for the second aspect, survival of comets in the classical zone of
the Oort cloud, the results depend on cluster parameters and details
of the modeling. \citet{letal2010}, assuming very low-mass clusters,
showed that the Oort cloud may capture extra-solar planetesimals quite
efficiently. It was not clear, though, whether the same model could
emplace the right number of objects into the fossilized inner zone of
the Oort cloud and thus explain the Sednoid population. Other studies
of more massive clusters generally did not reach the same level of
sophistication as the work of Levison et al.\ The investigations of
more massive clusters focus on the disruptive role of stellar
encounters with the classical Oort cloud
\citep[e.g.,][]{kq2008,nord2017}

We neglect the effects of the birth cluster on the formation of the
Oort cloud. Formation of the Oort cloud might have been a two-stage
process \citep[also see][]{brasser2008,bm2013,nord2017}. The first
phase involved dynamics in the birth cluster. This might have stored
bodies in the fossilized inner Oort cloud and left some population of
comets in the classical Oort cloud zone. Assuming that the Sun left
the cluster prior to the planetary instability, our model describes
what happened later on.
\smallskip

\noindent{\it Solar migration in the Galaxy.-- } Another badly
constrained issue of Oort cloud formation has to do with the solar
orbit in the Galaxy. This is because the Oort cloud was principally
built some $4$~Gyr ago \citep[e.g.,][]{dones2004}. However, there is
no exact constraint on the Sun's location in the Galaxy at that
epoch. Our model assumes the current orbit at all times, but very
likely the Sun performed a more complicated journey in our Galaxy
throughout its history. The most interesting aspect is its possible
radial migration \citep[see, e.g.,][]{Roskar2008,
Martinez-Barbosa2015, Frankel2018}. Migration would have directly
affected both galactic tide parameters and the frequency of stellar
encounters.

Several groups have studied Oort cloud formation in different galactic
environments \citep[e.g.,][]{bhk2010,krq2011,mbetal2017,hanse2018},
indicating that if the Sun was at a small galactocentric distance
during its early history, the effects would be somewhat similar to the
birth cluster. In particular, stronger tides and fiercer stellar
encounters would lead to the formation of the Oort cloud closer to the
Sun, extending its innermost zone perhaps near the Sednoid region. For
that to work, one should prefer models in which the Sun spent its
infancy at a rather small distance from the center of the
Galaxy. Additionally, a later solar excursion into this zone may cause
stronger erosion of the outer Oort cloud region which currently
provides observable LPCs. These accelerated losses may be somewhat
compensated by transfer from the inner regions of the Oort cloud
\citep[e.g.,][]{krq2011}.

With this perspective, we should consider our model a baseline before
we consider more complex possibilities. If future observations of
large-perihelion LPCs indicate a large mismatch with our predictions,
more careful studies involving models of the birth cluster and/or
solar radial migration in the Galaxy will be needed.
\smallskip

\noindent{\it Massive perturbers in the outer Solar system (planet
  9).-- } Several groups of researchers have recently suggested the
existence of a massive ($\approx 5$--20 Earth mass) body (planet 9)
roaming in the region beyond the classical Kuiper belt 
\citep[e.g.,][]{ts2014,bb2016,Batygin2019}. This body
was needed, according to them, to explain the non-uniform
distributions of secular angles (node and perihelion longitudes) of
about a dozen trans-Neptunian objects with extremely distant orbits
(i.e., $a> 150$~au, $q>35$~au). Planet 9 may also act as a perturber
that tilted the giant planets' invariant plane from the solar spin
direction \citep[e.g.,][]{bailey,lai,gomes2017} and produce
high-inclination, large-semimajor axis Centaurs
\citep[e.g.,][]{gomes2015,bb16b,Batygin2019}. While intriguing in many respects,
the hypothesis of the distant planet 9 is still debated. For instance,
analysis of observations by the Outer Solar System Origins Survey
(OSSOS), currently the most prolific survey of the trans-Neptunian
region, are still compatible with a uniform distribution of orbital
angles of distant objects when biases are properly accounted for
\citep[e.g.,][]{ossos6,ossos7}, although the originators of the 
planet~9 hypothesis find that the clustering is highly significant 
\citep{Brown2019}. The solar tilt may have been produced in an 
earlier phase of Solar system evolution
\citep[e.g.,][]{heller1993,thies2005,Batygin2019}, and in spite of search
campaigns, planet 9 still escapes direct detection.

As for the relation to cometary studies, \citet{nesvornySPC} examined
the role of planet 9 with the parameters originally suggested by
\citet{bb2016} for orbital and population characteristics of
short-period comets. They found that existence of planet 9 on this
orbit, with a mass of 15 Earth masses, makes it difficult to explain
the tight inclination distribution of Jupiter-family comets. This is
because planet 9 directly affects the properties of planetesimals in
the scattered disk, which acts as an immediate source for these
comets. As to the Halley-type comets, which are generally thought to
originate for the most part from the Oort cloud, \citet{nesvornySPC}
did not find any improvements to the model. In fact, when planet 9 was
taken into account, the match of the orbital elements of Halley-type
comets was not as good. Also, perturbations from planet 9 were
not found to significantly increase the flux of Halley-type comets
when compared to the model where only the galactic forces were taken
into account. Since LPCs originate from the Oort cloud, it is hard to
imagine that planet 9 would significantly improve the modeling of the
currently observed population of these comets. Future work may test
the effect of planet 9 on a putative population of LPCs with distant
perihelia.

With this experience, and because the current situation of planet 9 is
rather confused, we opted not to include it in the present study.

\smallskip

\noindent{\it Nongravitational accelerations in cometary dynamics.-- }
The original orbital elements inferred for comets, in particular their
original semimajor axes, depend on their levels of activity. So
whenever enough astrometric observations are available, orbit fitters
typically include nongravitational effects.  This procedure was
started and tested by the founders of MWC08
\citep[e.g.,][]{msy1973,ms1973,marsden1978}, and later on verified and
incorporated into the Kr\'olikowska et~al.\ catalogs
\citep[e.g.,][]{DK2011,KD2013,ketal2014}. As a rule of thumb, these
authors found that many apparently hyperbolic solutions among the
original orbits are moved to the category of very weakly-bound, but
elliptical solutions (often in the Oort peak). This is a very
interesting result, pointing to the importance of nongravitational
accelerations in cometary dynamics.

\citet{WT99} (see their Fig.~20) also noted that the predicted
distribution of the original semimajor axis changed when
nongravitational effects were included. They found that (i) the
dynamical effects correlate with the fading law, and (ii) simple
parametrization of the nongravitational effects worsens agreement with
the observations for comets on returning orbits with small semimajor
axes (perhaps because modeling of the recoil effects due to comet
activity is too simplistic).  As a result, while admitting their
importance, we also neglect nongravitational effects in our work.


\section{Results}\label{results}

\subsection{Properties of the Oort cloud}\label{our_cloud}
First, we take a brief look at the Oort cloud structure at the end of
our simulations, namely at $4.5$~Gyr. Since all our runs provide very
similar results, we use {\tt C1V1} as an example. We also note that
the situation becomes nearly stationary during the last Gyr, so our
analysis is representative of any moment, except for rare comet
showers, during that interval of time. As mentioned above, the Oort
cloud population declined in the {\tt C1V1} run by only $\simeq 12$\%
from $3.5$~Gyr to $4.5$~Gyr.

\begin{figure}[t!]
\epsscale{1.1}
\plotone{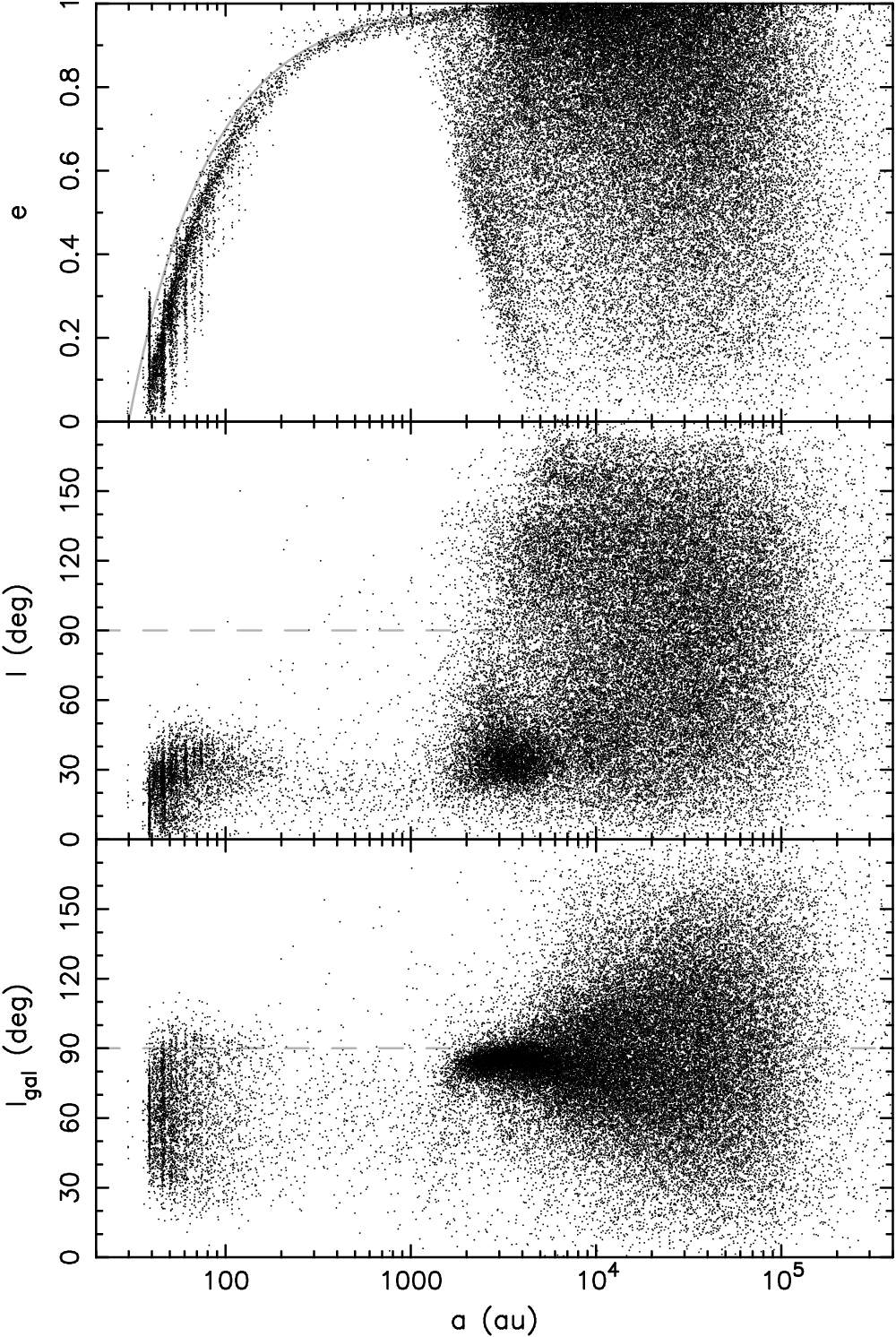}
\caption{Orbits of all $\simeq 51000$ particles remaining in the
 simulation {\tt C1V1} at $4.5$~Gyr: (i) semimajor axis
 vs.\ eccentricity (top), (ii) semimajor axis vs.\ inclination with
 respect to the ecliptic plane (middle), and (iii) semimajor axis
 vs.\ inclination with respect to the galactic plane (bottom). The
 gray line at the top denotes $q=30$~au (Neptune's heliocentric
 distance); the dashed lines in the middle and bottom panels denote
 polar orbits with $i=90^\circ$ (or $i_{\rm gal}=90^\circ$). The
 scattering disk (active, detached and the outer resonant
 populations) contains some $4600$ particles up to semimajor axis $a
 \simeq 1500$~au (with a majority of $\simeq 85$\% with $a < 200$~au.
 The inner and outer parts of the Oort cloud contain $\simeq 21500$
 and $\simeq 24300$ particles with $a < 15000$~au and $a > 15000$~au,
 respectively; 15000~au approximately represents the division between
 the non-isotropic and isotropic portions of the populations.}
 \label{oort_aei}
\end{figure}

Figure~\ref{oort_aei} shows the orbits of slightly more than $51000$
particles remaining in the {\tt C1V1} simulation at $4.5$~Gyr. The
innermost structures, with $a\lesssim 1000$~au, are of lesser
importance for our current work. They include the dynamically hot
classical Kuiper belt, resonant populations (including Plutinos), and
objects stored in the scattering disk, most with $q<35$~au. Only
objects interacting with high-order exterior resonances with Neptune
may become detached beyond this perihelion distance by processes
described in \citet{nvr2016} and \citet{ks2016}. The scattering
population is relevant to our study by constituting a pathway which
objects take to reach larger heliocentric distances. There is also a
population of a few objects with $q<30$~au and $a<1000$~au seen in
Fig.~\ref{oort_aei}. One would classify them as an extreme Centaur
population, which will further evolve toward short-period comets.
Some of these objects may also be considered in our analysis below as
returning long-period comets (unless they already performed so many
returns that they would be classified as faded objects). Large
surviving comets in this region continue their evolution towards the
class of Halley-type comets \citep[e.g.,][]{nesvornySPC}.
\begin{figure}[t!]
\epsscale{1.1}
\plotone{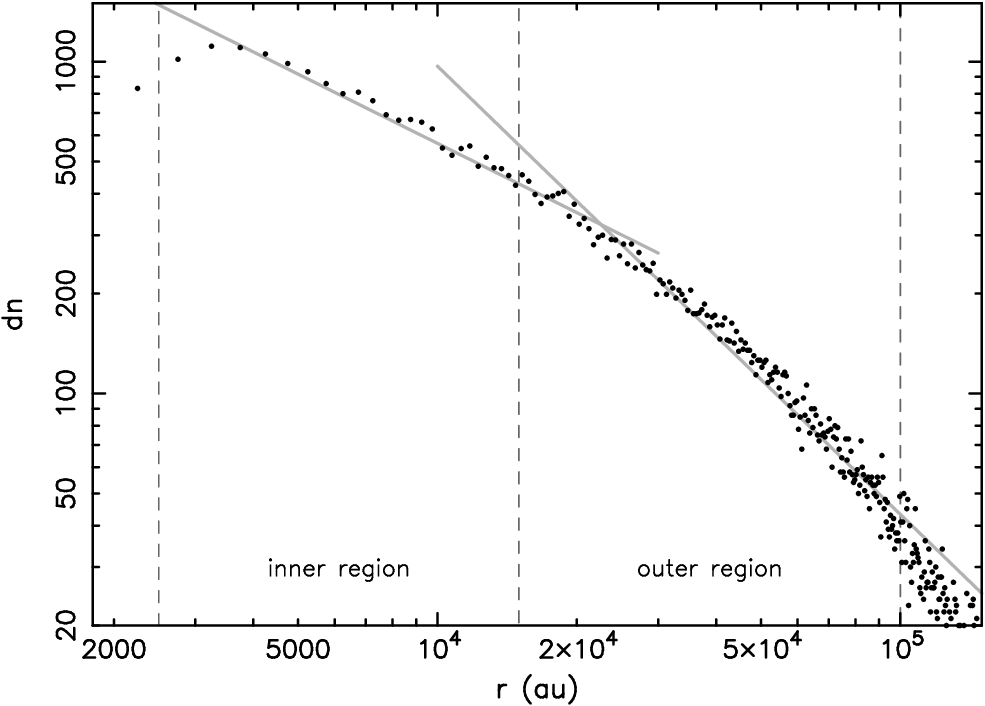}
\caption{Radial distribution of particles remaining in the simulation
 {\tt C1V1} at $4.5$~Gyr: the symbols give the number of particles
 $dn$ in bins $dr=500$~au (for the distribution law $dn=n(r)\,dr$).
 The dashed vertical lines roughly delimit the inner and outer parts
 of the cloud, where the power-law approximations $n(r)\propto
 r^{-\alpha}$ have different exponents: (i) $\alpha\simeq 0.72$ in
 the inner part, and (ii) $\alpha\simeq 1.35$ in the outer
 part. Below $r\simeq 2000$~au the population abruptly drops
 (also see Fig.~\ref{oort_aei}).}
 \label{oort_r}
\end{figure}
\begin{figure*}[t!]
\epsscale{1.1}
\plotone{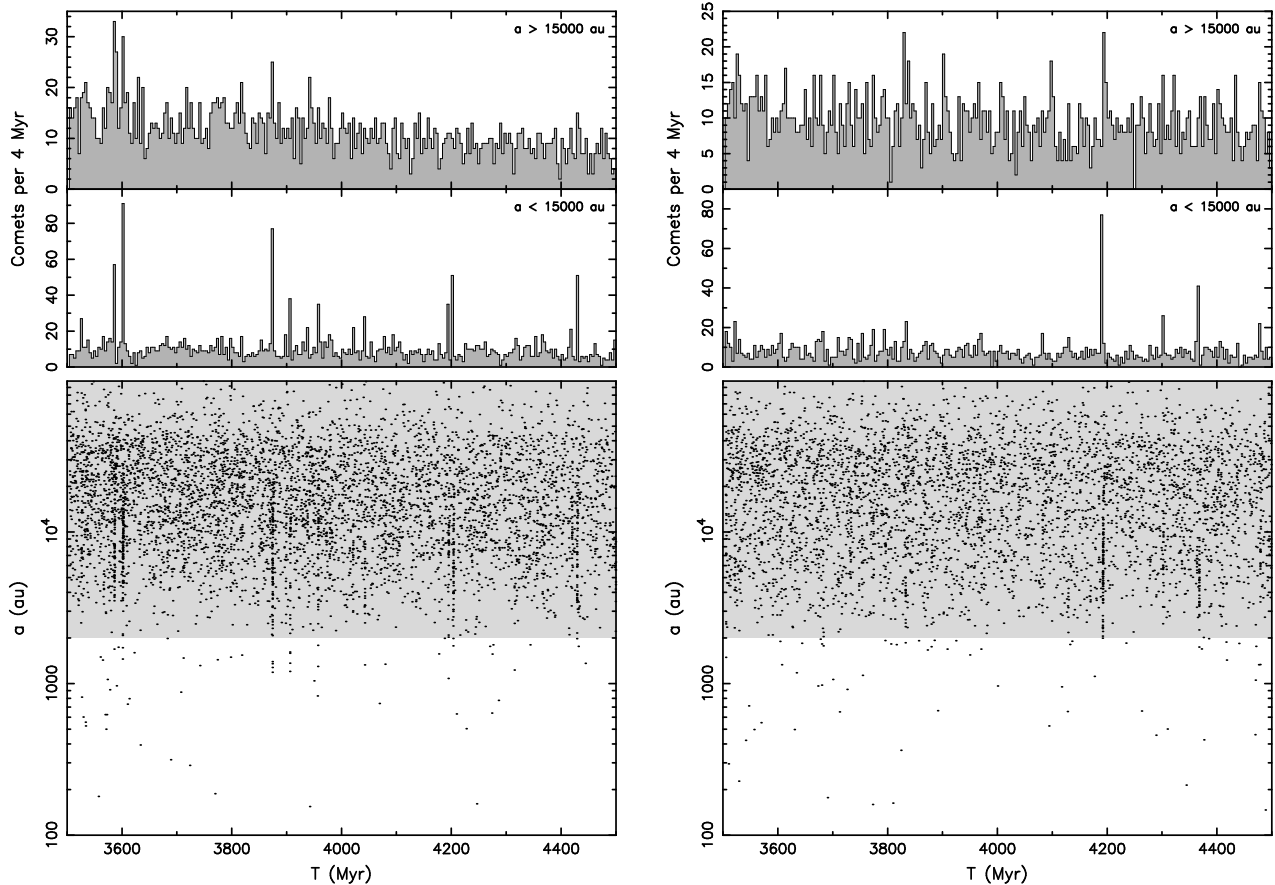}
\caption{Orbits of comets at their first appearances inside the
 heliocentric target zone $r\leq 20$~au: left panels for the {\tt
 C1V1} simulation, right panels for the {\tt C1V2} simulation. The
 abscissa is time in the last Gyr of the simulations. The bottom
 panels show the original semimajor axes $a$ of the orbits (the gray
 zone approximately delimits the heliocentric distance range of the
 Oort cloud). The upper panels show the number of new comets in
 $4$~Myr bins. We separate cases originating from the outer/inner
 parts of the Oort cloud (with $a>15000$~au and $a<15000$~au,
 respectively). Cometary showers, associated with particularly strong
 stellar encounters (Fig.~\ref{nstar}), are clearly seen coming from
 the inner Oort cloud, while their signal is absent in the outer Oort
 cloud.}
 \label{nc_20}
\end{figure*}

Further on, at $a\gtrsim 1500$~au, we reach the realm of the Oort
cloud. The lower two panels in Fig.~\ref{oort_aei}, showing the
inclination with respect to the ecliptic (middle) and galactic
(bottom) planes, best illustrate the two distinct regions, the inner
and outer Oort clouds. The anisotropic nature of the inner part, from
semimajor axes $\simeq 1500$~au to $\simeq 15000$~au, is readily
explained by the orbital evolution due to galactic tides
\citep[e.g.,][]{hetal2007,hk2015,fetalIca2017}.  In this region the
tides are too weak, such that orbits pulled from the tail of the
scattered disk perform less than one cycle of their secular evolution
\citep[see, e.g., Fig. 3 in][]{fetalIca2017}. The slow evolution
towards small eccentricity values produces the visible edge of the
inner Oort cloud and also implies that inclinations with respect to
the galactic plane are strongly concentrated towards $90^\circ$, where
the secular evolution spends most of the time
\citep[e.g.,][]{hetal2007}. Because the mean inclination of the
scattered disk is $\simeq 60^\circ$ in this reference frame, the
orbits do not overcome the $90^\circ$ limit in the quadrupole tidal
model. They may scatter over this limit only by occasional tugs due to
passing stars. Transformed to the ecliptic frame, this concentration
occurs at $\simeq 35^\circ$, with a weaker concentration near $\simeq
150^\circ$. In the outer part of the Oort cloud, beyond semimajor axes
$\simeq 15000$~au, the inclination distribution becomes nearly
isotropic in space. Orbits in this region have performed at least
several secular cycles due to the tides, helping in their mixing. More
importantly, beyond about $\simeq 40000$~au the purely secular model
is not justified, because the strength with which orbits are bound to
the Sun becomes similar to the tidal effects. The orbits become
essentially chaotic \citep[also see][]{brasser2001}. Finally, orbital
mixing due to the stellar passages becomes a vigorous process in this
zone.
\begin{figure*}[t!]
\epsscale{1.1}
\plotone{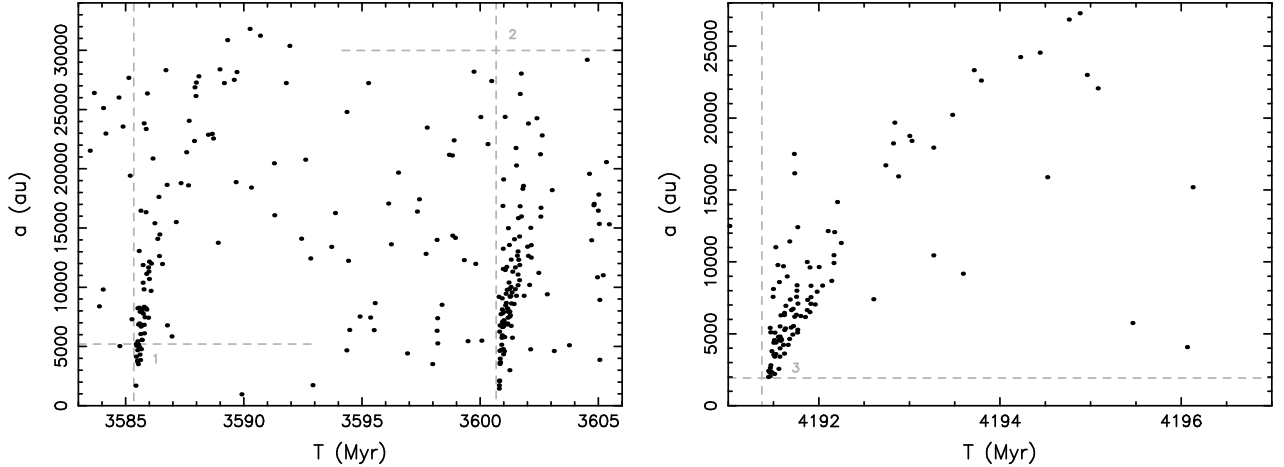}
\caption{Two examples of comet showers into the heliocentric target
 zone $r\leq 20$~au (highly time-zoomed data from the bottom panels
 of Fig.~\ref{nc_20}).  Orbits of comets at their first appearance,
 with time as abscissa and original semimajor axis as ordinate, for
 the stellar encounters highlighted by labels 1, 2 and 3 of
 Fig.~\ref{nstar} are shown. The horizontal dashed lines indicate the
 smallest distance at which the star encounters the Sun, while the
 vertical dashed lines indicate the epochs of the closest
 encounters. Left: Two encounters, fortuitously close in time, from
 simulation {\tt C1V1}. The first, labeled 1, represents a star of
 $0.78$~M$_\odot$ mass encountering the Solar system with an
 asymptotic velocity of 24~km s$^{-1}$ , while the second, labeled 2,
 represents a star of $9$~M$_\odot$ mass encountering the Solar
 system at $14$~km s$^{-1}$. Right: a single encounter in the
 simulation {\tt C1V2} with the largest recorded value $N_{\rm
 par}\simeq 540$, corresponding to a $0.26$~M$_\odot$ star
 encountering the Solar system at $16$~km s$^{-1}$ at a minimum
 distance of $\simeq 1920$~au. The ``triangular'' shapes of the
 showers imply that comets from larger-$a$ orbits generally arrive
 slightly later, as expected.}
 \label{shower}
\end{figure*}
\begin{figure*}[t!]
\epsscale{1.1}
\plotone{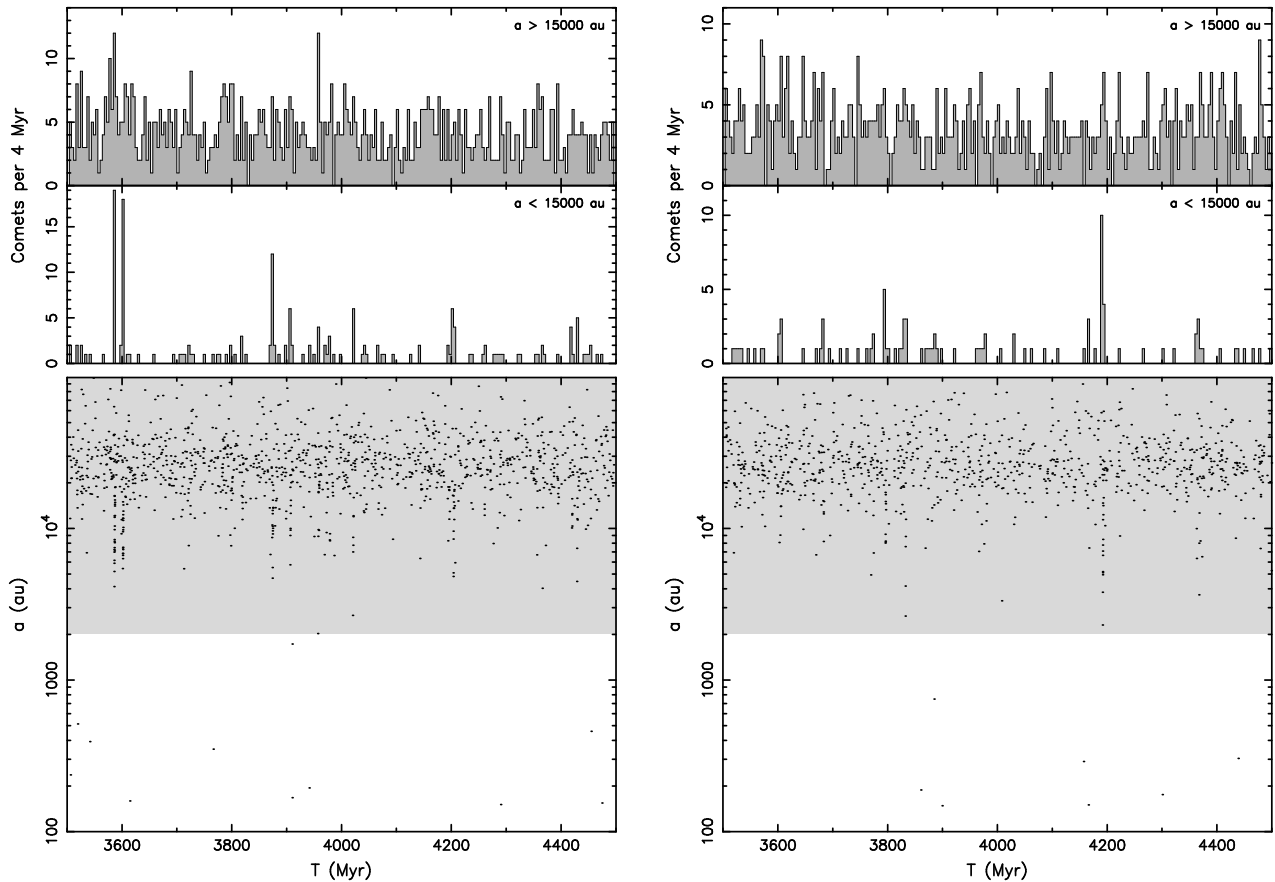}
\caption{The same as in Fig.~\ref{nc_20}, but now for the first
 appearance of comets inside a much more restricted heliocentric
 target zone $r\leq 5$~au.  Now the inner Oort cloud is basically
 absent as an apparent source of new comets except during strong
 cometary showers. Most of the flux of new comets now comes from the
 outer Oort cloud with $a>15000$~au.}
 \label{nc_5}
\end{figure*}

Figure~\ref{oort_r} shows the radial heliocentric distribution of
comets in the Oort cloud at the end of our simulation {\tt C1V1}. We
plot the number of objects $dn$ in uniform radial steps
$dr=500$~au. While not exactly a power law, the incremental
distribution function $n(r)=dn/dr$ may be in parts approximated with
$n(r)\propto r^{-\alpha}$. In the inner cloud, we find $\alpha\simeq
0.72$, while in the outer cloud, we find $\alpha\simeq 1.35$,
steepening to a thermalized value of $1.5$ at the very outer edge of
the cloud (beyond $\simeq 50000$~au). [Note our $\alpha = 1.5$ corresponds 
to $\alpha = -3.5$ as defined by \cite{Duncan1987AJ}]. The population of 
the inner region of the cloud is comparable, but actually slightly smaller 
than, that of the outer region. This is also related to the shallow
power-law exponent (inspecting our other simulations, we have $\alpha$
always in the range of $0.68$ to $0.77$ in the inner Oort cloud). The
Oort cloud formed in our model therefore has a less populous inner
region, if compared to some previous models (often assuming the
thermal exponent $1.5$ extending throughout the whole cloud). However,
the results here are comparable to several other models such as
\citet{dones2004}.  Note that the Oort cloud fills in from the outer
parts to the inner zone. Therefore, details of the population in the
inner cloud depend sensitively on the late deposition of planetesimals
in the tail of the scattered disk in the migration scenario. We find
that the inner zone starts to fill effectively at $\simeq 250-300$~Myr
\citep[compare with Fig.~8 of][]{dones2004}. This explains why our
{\tt C1} and {\tt C2} models (see Table~\ref{tab1}) produce
rather comparable results: the assumed timescales $\tau_1$ and
$\tau_2$ are still short, if compared to the inner Oort cloud
filling timescale.

\subsection{Comets at their first appearance}\label{res_1}
As discussed by \citet[][their class V$_1$]{WT99}, the properties of
LPCs at their first appearance in the target zone may be a useful
starting point for their analysis as a whole. We start with discussion
of the orbital parameters of new comets in the largest target zone
monitored during the last Gyr in our simulations, namely the
heliocentric sphere of $20$~au (Sec.~\ref{out}). This zone is larger
than the currently observable region, but future observations hope to
reach this zone.  While speaking about new comets here, we point out
that we do not know their orbital evolution before appearing in this
target zone. In particular, we do not know whether a particular orbit
jumped in from a very distant-perihelion state, or whether its
perihelion was slowly evolving towards the $20$~au limit.

In Figure~\ref{nc_20}, the lower panels show the original semimajor
axes $a$ of newly-appeared LPCs in the $r\leq 20$~au zone during the
last Gyr in the {\tt C1} simulations. In the two variants of stellar
encounters, {\tt V1} and {\tt V2}, there are about $5500$ and $4350$
data points in the respective runs.  Two patterns are seen: (i)
randomly distributed data with no strong correlation between time and
original semimajor axis, and (ii) occasional sequences of new comets
strongly localized in time. The former is a background population,
originating from the entire Oort cloud. In each of the two variants,
comparable numbers of comets arrive from the inner and outer parts of
the Oort cloud, roughly in proportion to their populations
(Fig.~\ref{oort_r}). The second, time-correlated component in the
population of new comets constitutes showers after the most important
stellar encounters. Their occurrence coincides very well with the
events for which $N_{\rm par}\geq 40$ in Fig.~\ref{nstar}. Note that
cometary orbits in these showers apparently originate only from the
inner part of the Oort cloud, for which $a\lesssim 15000$~au. This is
again well documented in the upper panels of Fig.~\ref{nc_20}, where
we show the number of comets collected in $4$~Myr wide bins in
time. The dominance of the inner Oort cloud in its contribution to the
shower periods is well known from previous studies
\citep[e.g.,][]{hetal1987,h1990,fetalIca2011}, though that work often
focused on smaller heliocentric target zones. The largest contrast
between the number of new comets in the modeled showers and the
long-term mean of the background signal is $\simeq 4-5$. This is
in accord with results of \citet{fetalAA2011} and \citet{fetalIca2011}.

Figure~\ref{shower} provides a zoom of the lower panels in
Fig.~\ref{nc_20} for three prominent showers: (i) the left panel
illustrates the results of the two stellar encounters labeled 1 and 2
in Fig.~\ref{nstar} from the {\tt C1V1} simulation, while (ii) the
right panel illustrates comets from the strongest stellar encounter,
labeled 3 in Fig.~\ref{nstar}, from the {\tt C1V2} simulation
(parameters of the stellar trajectories relative to the Sun are given
in the caption). Clearly, comets having smaller-$a$ orbits
statistically arrive first, because of their smaller orbital
periods. However, because the encounters occur at a random phase of
the orbital motion of the comet (i.e., some comets with large
semimajor axes are near perihelion), some with larger-$a$ orbits may
also arrive nearly instantly. Nevertheless, those which are delayed
with respect to the stellar passage must also arrive from very wide
orbits. This produces the triangular shape of the region where the
shower comets are concentrated. It has been noted in several earlier
studies that the strongest showers are not necessarily produced by
encounters with the most massive stars.  There are very rare and may
happen only once or twice in the history of the Solar system (for
stars with mass $\gtrsim 10$~M$_\odot$, say). Statistically more
important are very close, and low-velocity, encounters with sub-solar
mass stars. These may happen once per $\simeq 100-150$~Myr, on average
\citep[e.g.,][]{hetal1987,h1990,fetalIca2011}.

Figure~\ref{nc_5} shows the same information as Fig.~\ref{nc_20}, but
now restricted to the heliocentric target zone $r\leq 5$~au. This
range is now compatible with the perihelia of the presently observed
comets (Sec.~\ref{obs}).  As expected, almost all members in the
background population of new comets come only from the outer part of
the Oort cloud. With a few outliers, the inner Oort cloud becomes
active only during the strongest comet showers
\citep[see][]{hills1981,hetal1987,h1990}. This is expected, because
tides are efficient enough to fill the phase space region of
LPC orbits reaching $q < 5$~au (their ``loss cone'') only for $a\gtrsim
30000$~au. Comets with $a$ down to about
$15000$~au may also contribute, if they creep their perihelia through
the planetary zone above the orbit of Saturn and eventually increase
their semimajor axes enough by planetary perturbations before the
final jump into the observable zone \citep[e.g.,][]{kq2009}.  However,
orbits with semimajor axes in the inner Oort cloud undergo changes in
perihelion distance in one orbit that are too small, so that Jupiter
and Saturn efficiently eliminate them before they can appear with
perihelia within Jupiter's orbit \citep[see also][]{retal2008,fetalIca2011,fetal2014}.
\begin{figure*}[t!]
\epsscale{1.1}
\plotone{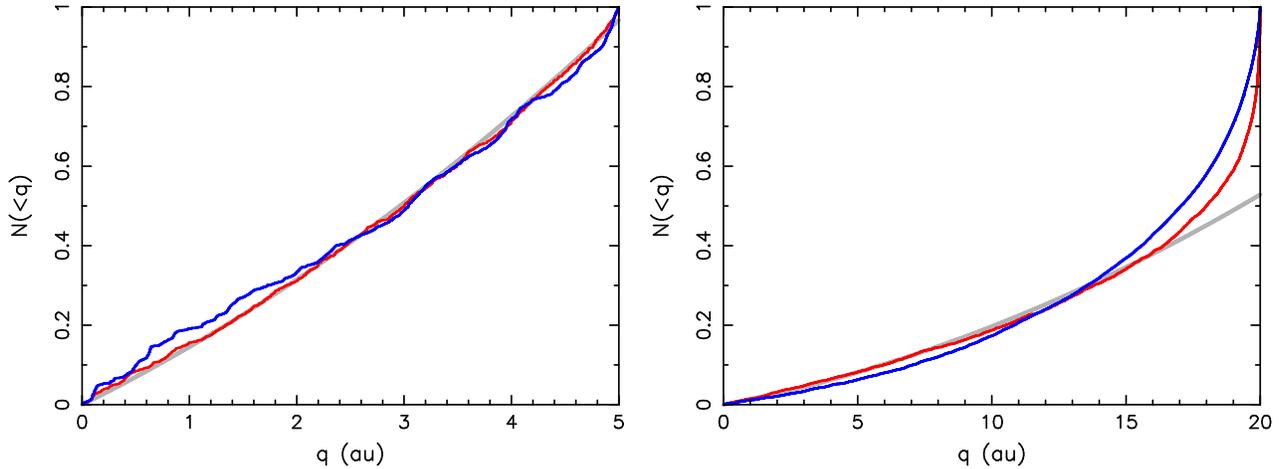}
\caption{Cumulative distribution $N(<q)$ of LPCs' perihelia $q$ in two
 different heliocentric target zones: (i) $q\leq 5$~au (left), and
 (ii) $q\leq 20$~au (right), both in the simulation {\tt C1V1} (data
 during the strongest comet showers with $N_{\rm par}\geq 40$ were
 eliminated from the distributions shown). The red curves are for
 comets at their first appearance in the target zone. The gray curves
 are linear-quadratic fits of the new comet $q$ distributions, namely
 (i) $N(<q)\propto q + 0.09\,q^2$ in the left panel, and (ii)
 $N(<q)\propto q + 0.06\,q^2$ in the right panel. The latter matches
 the $N(<q)$ distribution sufficiently well only until $q\simeq
 15$~au, beyond which the number of new comets steeply rises. The
 blue curves show the cumulative distributions $N(<q)$ when all
 returning comets are included (no physical fading).}
 \label{nc_qdist}
\end{figure*}
\begin{figure}[t!]
\epsscale{1.1}
\plotone{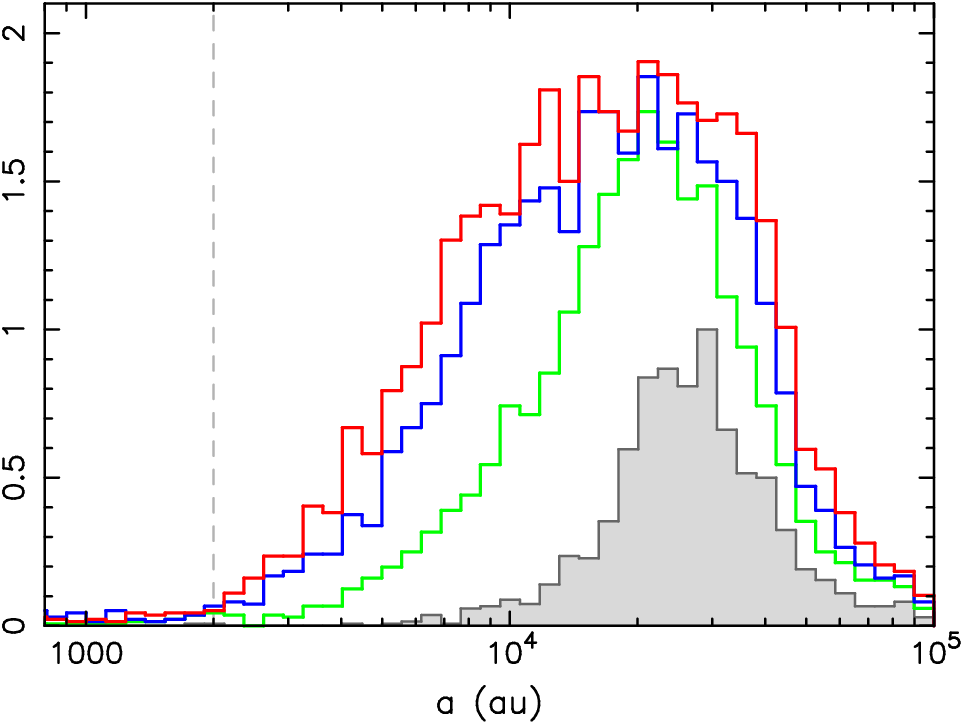}
\caption{Distribution of semimajor axes of LPCs at their first
 appearances in heliocentric zones of different radius $r$ in
 simulation {\tt C1V1} (data during the strongest comet showers with
 $N_{\rm par}\geq 40$ were eliminated from the distributions shown).
 The gray histogram is for $r\leq 5$~au and roughly corresponds to
 the currently observed population (see Figs.~\ref{mw_a}, \ref{w_a}
 and \ref{w_qi}); this is the ``traditional'' Oort peak of new
 comets. The color histograms show the same, but for larger
 heliocentric target zones: (i) $r\leq 10$~au (green), (ii) $r\leq
 15$~au (blue), and (iii) $r\leq 20$~au (red). The ordinate is
 arbitrarily normalized to unity for the maximum of the $r\leq 5$~au
 histogram.  The total number of new comets in the $r\leq
 20/15/10$~au zones is $\simeq 4.5/3.9/2.7$ times larger than the
 population entering the $r\leq 5$~au zone.}
 \label{nc_ahist}
\end{figure}

The results discussed above confirm the critical role of the radius
$r$ of the target zone around the Sun where new comets are being
recorded, especially if crossing the Jupiter-Saturn zone. We repeated
our analysis for several choices of $r$.  Focusing on the background
population of new comets, each time we eliminated comets in the
strongest showers (stellar encounters with $N_{\rm par}\geq 40$) from
the data.  Figure~\ref{nc_ahist} shows the incremental distribution of
the original semimajor axes of LPCs as they first arrive in the target
zone during the last Gyr of our simulation {\tt C1V1} (results for
other simulations are very similar). The gray distribution corresponds
to the data in Fig.~\ref{nc_5}, thus $q\leq 5$~au comets.  This is the
classical Oort peak of nearly parabolic comets seen in the observed
population of LPCs (see Fig.~\ref{mw_a}). When extending the limiting
$r$ to larger values (green to red curves in Fig.~\ref{nc_ahist}), we
note two systematic effects: (i) original orbits with smaller $a$
values start to dominate and the overall distribution of $a$ becomes
broader, and (ii) the total population of new comets increases
approximately proportionally to $r$.  This is because the inner Oort
cloud is now able to contribute to the population of new comets
\citep[see also][]{st2016,fetalAA2017}. 

We have not yet discussed the distribution of the new comets'
perihelia. This information is shown in Fig.~\ref{nc_qdist} for two
heliocentric target zones, $r\leq 5$~au on the left and $r\leq 20$~au
on the right. Focusing first on the restricted $5$~au heliocentric
zone (left panel), we confirm results from previous studies
\citep[e.g.][]{WT99,fetalAA2017} that the cumulative perihelion
distribution of new comets is very well fitted with a linear term and
a small quadratic contribution. Interestingly, if we use the orbits of
both new and all returning comets before their dynamical elimination
(therefore applying no fading), the perihelion distribution is not
changed much (blue curve in the left panel of Fig.~\ref{nc_qdist};
therefore, it is to be expected that even when applying some fading
law the perihelion distribution would still behave the same). The top
two panels in Fig.~\ref{c_orbits} may offer the explanation: once the
orbits happen to decrease their perihelia below Jupiter's orbit, its
value stays approximately constant until elimination. Obviously, a
possible caveat of our simulation is the absence of the terrestrial
planets. It is yet to be seen whether their gravitational
perturbations modify these perihelia distributions.

Extending the heliocentric target zone again to $20$~au, we obtain the
results shown in the right panel on Fig.~\ref{nc_qdist}. The
linear-quadratic trend in the cumulative perihelion distribution of
new comets (red curve) continues to about $q\simeq 15$~au.  Beyond
this point, the population of new comets increases steeply. A similar
behavior is also seen in the perihelion distribution of all new and
returning comets in this zone (blue curve), though the difference with
respect to the statistics of new comets is now a little larger. This
is because the returning comets have more space to random-walk their
perihelia, especially above the orbit of Saturn.
\begin{figure}[t!]
\epsscale{1.1}
\plotone{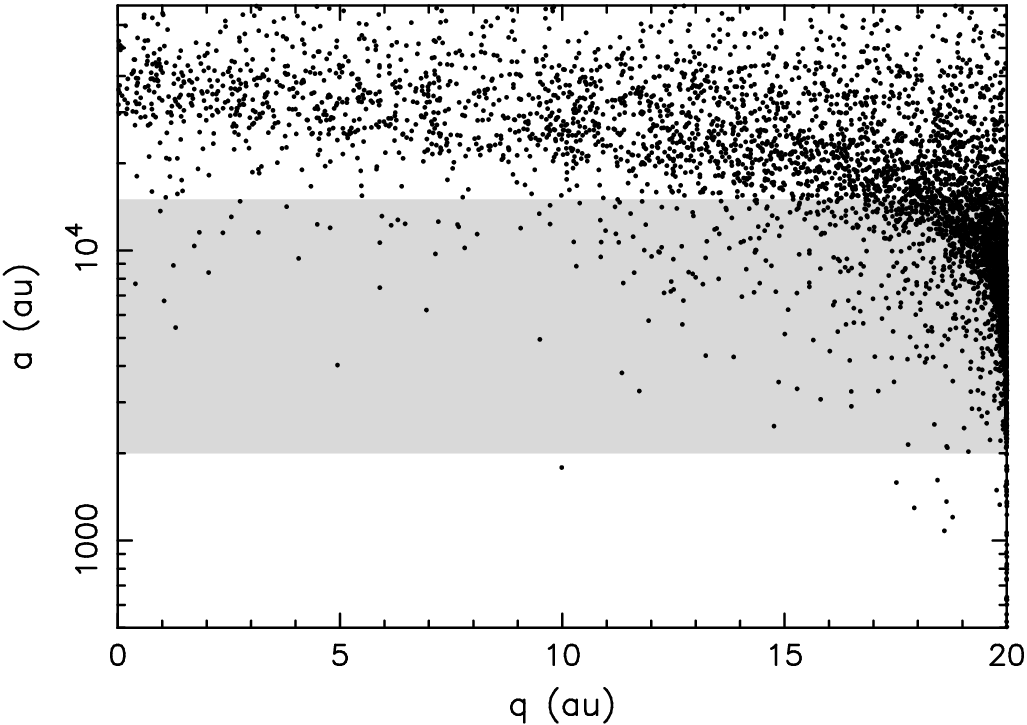}
\caption{Correlation between the perihelion distance $q$ (abscissa)
 and the original semimajor axis $a$ (ordinate) for LPCs at their
 first appearance in the heliocentric target zone $r\leq 20$~au. We
 used data from the last Gyr of simulation {\tt C1V1} used (the
 signal from the strongest comet showers with $N_{\rm par}\geq 40$
 was eliminated from this plot). The gray rectangle shows the
 approximate location of the inner part of the Oort cloud. At small
 heliocentric distances, up to $10-12$~au, the apparent source zone
 of new LPCs is in the outer Oort cloud. Beyond $\simeq 15$~au the
 inner Oort cloud starts to contribute, and near $20$~au the inner
 Oort cloud becomes the dominant source zone (also see
 Figs.~\ref{nc_20}, \ref{nc_5}, \ref{nc_ahist} and \ref{nc_qdist}).}
 \label{nc_qa}
\end{figure}
\begin{figure*}[t!]
\epsscale{1.1}
\plotone{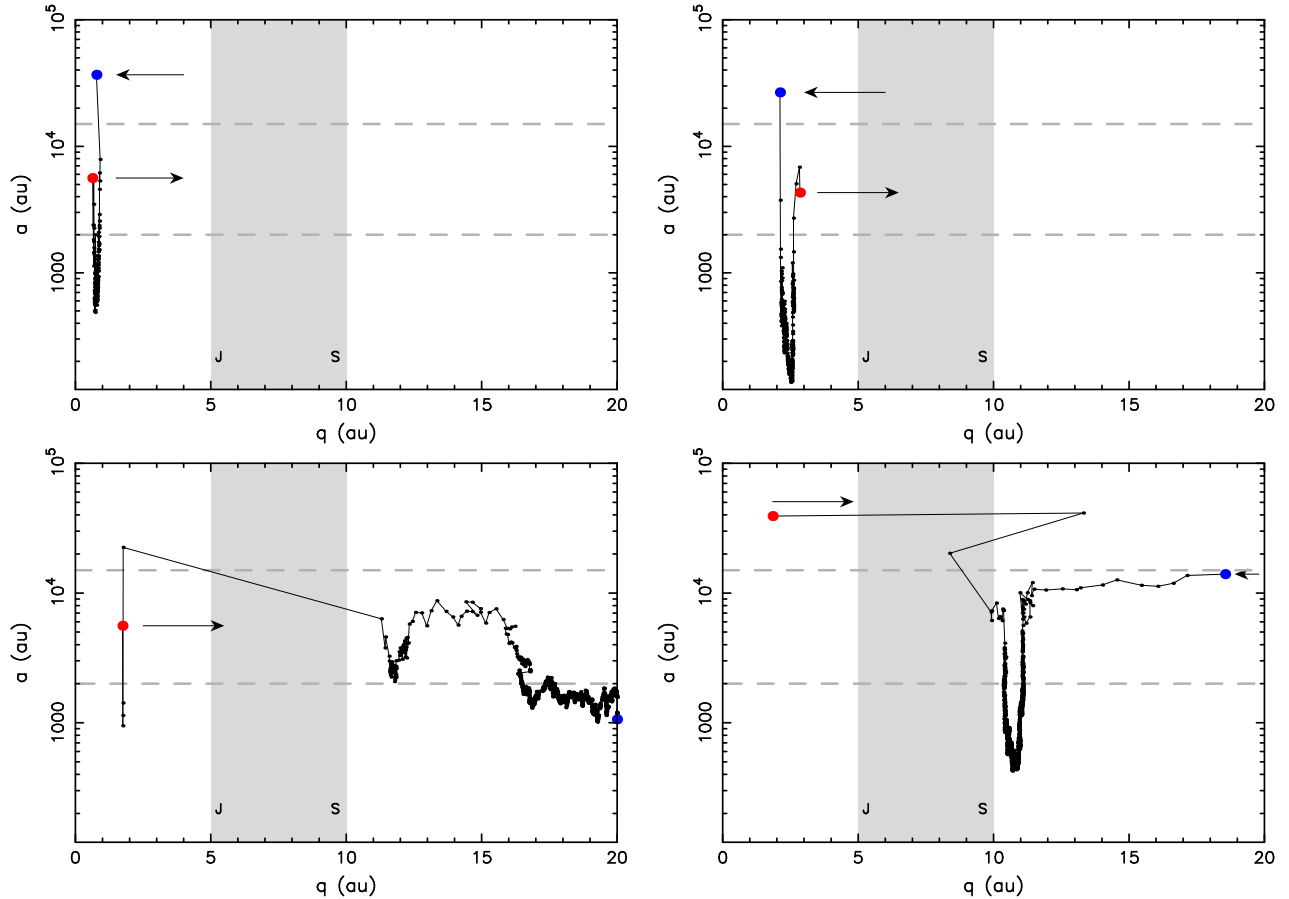}
\caption{Four examples of LPC orbital evolution in our simulation {\tt
 C1V1}: for each return we show the perihelion distance $q$
 (abscissa) vs original semimajor axis $a$ (ordinate). Data are
 collected only when $q\leq 20$~au (the orbital evolution before
 reaching this limit is not recorded). The first appearance in the
 $20$~au heliocentric zone is shown by the blue symbol, the last by
 the red symbol (no physical fading is assumed, so the comet is
 dynamically eliminated after its last return; arrows indicate
 the sense in which the orbits are injected or ejected from the
 monitored zone). Data for the
 intermediate returns are shown by black symbols and connected by
 lines to indicate their sequence. The gray rectangle shows the
 Jupiter-to-Saturn heliocentric zone, and the gray-dashed horizontal
 lines indicate the inner Oort cloud. The upper panels are examples
 of ``jumpers'', LPCs that appear in the currently observable
 $q\lesssim 5$~au zone without having a prior perihelion evolution
 closer than $20$~au.  The bottom panels are examples of
 ``creepers'', LPCs that had many tens of returns in the ice-giant
 region above Saturn's orbit before first appearing in the currently
 observable $q\lesssim 5$~au zone. Note that all these cases appear
 to be injected into the $q\lesssim 5$~au zone from the outer Oort
 cloud (i.e., initially with $a>15000$~au).}
 \label{c_orbits}
\end{figure*}

The rapid increase in the population of new comets beyond about
$15$~au is intriguing. To shed more light on this topic, we plot the
correlation between perihelia $q$ and original semimajor axes $a$ of
new comets using our outermost target zone of $r\leq 20$~au in
Fig.~\ref{nc_qa}. Orbits with $a\gtrsim 30000$~au populate all
perihelion distances about equally.  This is because the magnitude of the
perihelion change $\Delta q$ in one orbit, roughly expressed as $\Delta q\propto
\sqrt{q}\,a^{7/2}$ for both tidal effects and stellar perturbations
\cite[e.g.,][]{retal2008,rickman2010}, is large enough to decrease the
perihelion distance to arbitrarily small values from $q\geq 30$~au
initial orbits.  If the orientation of the comet's orbit is in a
certain range, the outer Oort cloud thus may contribute by injecting
comets to perihelia below Jupiter's orbit. As the target zone slightly
increases, the outer Oort cloud may contribute from slightly lower-$a$
initial orbits and this produces the small quadratic term in the
cumulative distributions in Fig.~\ref{nc_qdist}. Beyond $q\simeq
15$~au, the population of the inner Oort cloud also contributes. This
time a variety of orbital evolutions before entering the target zone
are possible. Either a direct jump or, more often, a gradual decrease
of perihelion in small steps (creeping) can occur. Neptune, with its
smaller mass than Jupiter or Saturn, is not a big obstacle to this
process. In this way, a significant number of comets may gradually
evolve to perihelia between 15 and 20~au from the inner Oort cloud.
This also implies that these distant-perihelia comets have orbits that
are not isotropic. Instead, their inclination distribution
approximately reflects the source zone, with an overabundance of
orbits with $\simeq 40^\circ$ inclination to the ecliptic plane.

In order to illustrate some of the principles mentioned above, and
bridge into the next section, in Fig.~\ref{c_orbits} we present a few
examples of orbital evolutions from our simulations (no physical
fading was included in these illustrations). The upper two panels show
a typical jumper evolution: the original semimajor axis in the outer
part of the Oort cloud allows a very large change in perihelion
distance, landing at $q\simeq 1$~au or $2$~au. Next, the perihelion
distance stays approximately constant, while the semimajor axis
drifts. This is the classical characteristic dynamics of the returning
population of LPCs, as we discuss in the next section. The two bottom
panels describe what has been characterized as creeping evolution
\citep[e.g.,][]{kq2009,fetal2014}. Thinking about currently observable
comets, both cases shown in the lower panels enter the $q<5$~au zone
from the outer Oort cloud (at least in terms of the original semimajor
axis). Yet, they experienced a significant perihelion evolution in the
Saturn--Uranus zone. At least the bottom left case might have
initially walked in from the inner Oort cloud. Prerequisite to this
evolution is a sufficient increase in semimajor axis before jumping
into the observable zone (to perform the necessary $\Delta q$ relative
to its instantaneous value). Related to our previous discussion of the
data shown in Fig.~\ref{nc_qa}, both of the orbits in the lower panel
enter the $20$~au target zone as new comets with very large perihelion
values ($\ge 18$~au), both representative of the inner Oort cloud
source. 
\begin{figure}[t!]
\epsscale{1.1}
\plotone{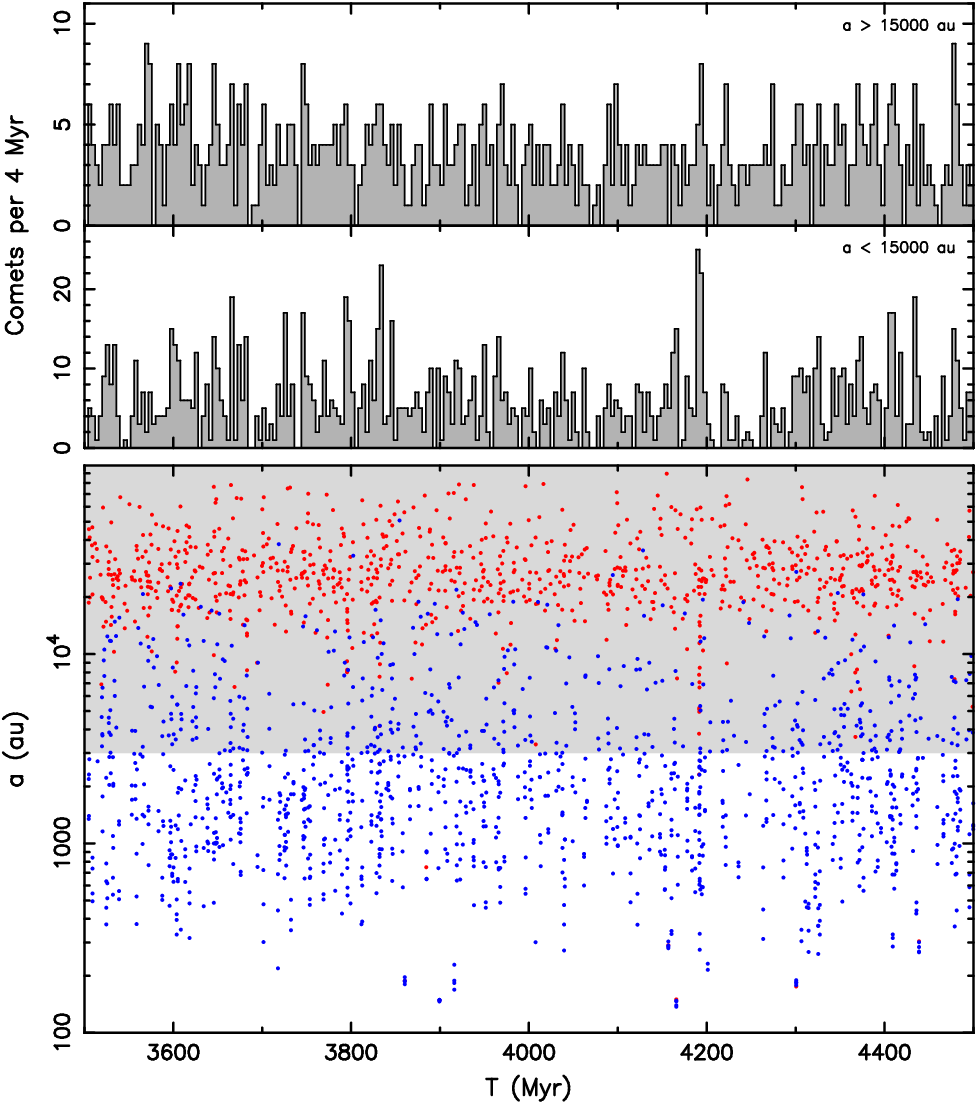}
\caption{The same as in the right panels of Fig.~\ref{nc_5}, but now
 new comets in the heliocentric target zone $r\leq 5$~au are also
 allowed to contribute by their subsequent returns (up to four of
 them). Data from the last Gyr of the simulation {\tt C1V2} are used
 here. The bottom panel shows the original semimajor axis $a$, with
 red symbols for new comets and blue symbols for returning
 comets. The upper panels show the number of comets in 4~Myr
 bins. We separate cases originating in the outer/inner parts of the
 Oort cloud (with $a>15000$~au and $a<15000$~au), respectively.}
 \label{nc_ret}
\end{figure}

\subsection{Returning population of comets}\label{res_all}
We now briefly demonstrate the effect of subsequent returns of
LPCs. Previous experience showed (see, e.g., Sec.~\ref{fad}) that many
observed LPCs do not survive a large number of returns. So, while in
our simulations some particles underwent hundreds of perihelion
passages before experiencing dynamical elimination, our preferred
fading law allows only a few returns before the typical comet
experiences physical elimination.

For the sake of illustration, we thus extended the data about new comets
in the {\tt C1V2} simulation from Fig.~\ref{nc_5} (right panels) by
allowing up to an additional four returns. We also maintain $5$~au as
the radius of the target zone, in which these comets are assumed to be
observed. The result is shown in Fig.~\ref{nc_ret}: the bottom part
shows the original semimajor axes of recorded comets, with red symbols
for their first appearances and blue symbols for their subsequent
returns. We note the following.

First, while occasionally the returning orbits have semimajor axes in
the outer Oort cloud, most often they are shifted to much smaller
values. These changes in semimajor axis are produced by planetary
perturbations. New comets, after first visiting the inner Solar
system, typically suffer a change in the inverse of their original
semimajor axis of order $\delta(1/a)\simeq (0.5-1)\times
10^{-3}$~au$^{-1}$ due to interactions with the giant planets
\citep[e.g.,][]{Everhart1968, rickman2010}. About half of them are
lost to interstellar space, while the other half is stabilized to much
more strongly bound orbits out of the Oort peak. They populate the
hump of returning orbits with semimajor axis values between a few
hundred and a few thousand au seen in the data compiled by MWC08
(Fig.~\ref{mw_a}). This is satisfactorily indicated by the set of blue
symbols in Fig.~\ref{nc_ret}.

The second observation concerns the significance of comet showers. The
top panel indicates that their visibility is further diminished if the
population of returning comets is added. In the case of new comets,
those with $a\leq 15000$~au were basically only shower members
(Fig.~\ref{nc_5}); now the returning component from the outer Oort
cloud feeds the background signal in this category of
orbits. Additionally, even the shower signal may get spread over a
longer interval of time when returns of their members are included in
the data. This is because the initial shower orbits may have orbital
periods ranging from a little less than $100$~kyr to several Myr and
this defocuses the narrow signal of the shower.

Finally, data in the top panels of Fig.~\ref{nc_ret} indicate that the
long-term mean flux of LPCs in this simulation, new and returning comets with
up to 4 returns, is $\simeq 1.5-2$~Myr$^{-1}$ from the initially integrated
million particles in the source zone. 
In the next Section we shall elaborate on the expected flux. We shall
also use the calibration of the planetesimal disk population to
express these data in terms of the mean size of the observed
long-period comets in our model.
\begin{figure*}[t!]
\epsscale{1.1}
\plotone{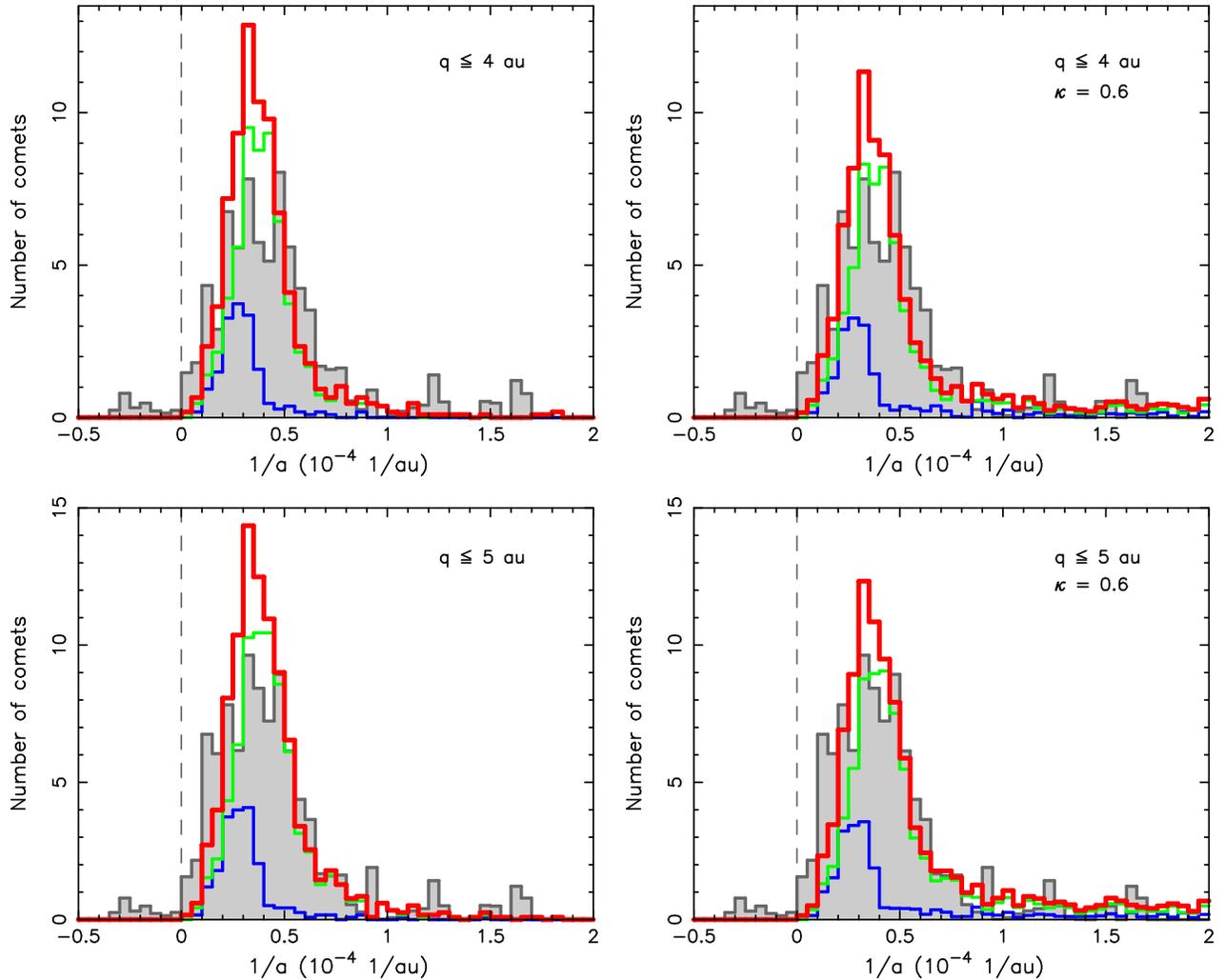}
\caption{Comparison between data (gray histogram) and simulations for
 LPCs on nearly parabolic orbits: incremental distribution of the
 inverse values $1/a$ of the original semimajor axis. The top panels
 are comets with $q\leq 4$~au, while the bottom panels are comets
 with $q\leq 5$~au. The panels on the left use only simulated new
 comets in the target zone, while the panels on the right also
 include returning comets with Whipple's power-law parametrization of
 the fading law and exponent $\kappa=0.6$ (Section~\ref{fad}). The
 red histogram is the total population predicted by our {\tt C1V1}
 simulation normalized to the same number of comets as the data ($78$
 in the upper panels, and $95$ in the bottom panels). The blue
 histogram indicates the jumper component and the green histogram
 shows the creeper component in all simulated comets.}
 \label{w_a_fit}
\end{figure*}
\begin{figure*}[t!]
\epsscale{1.1}
\plotone{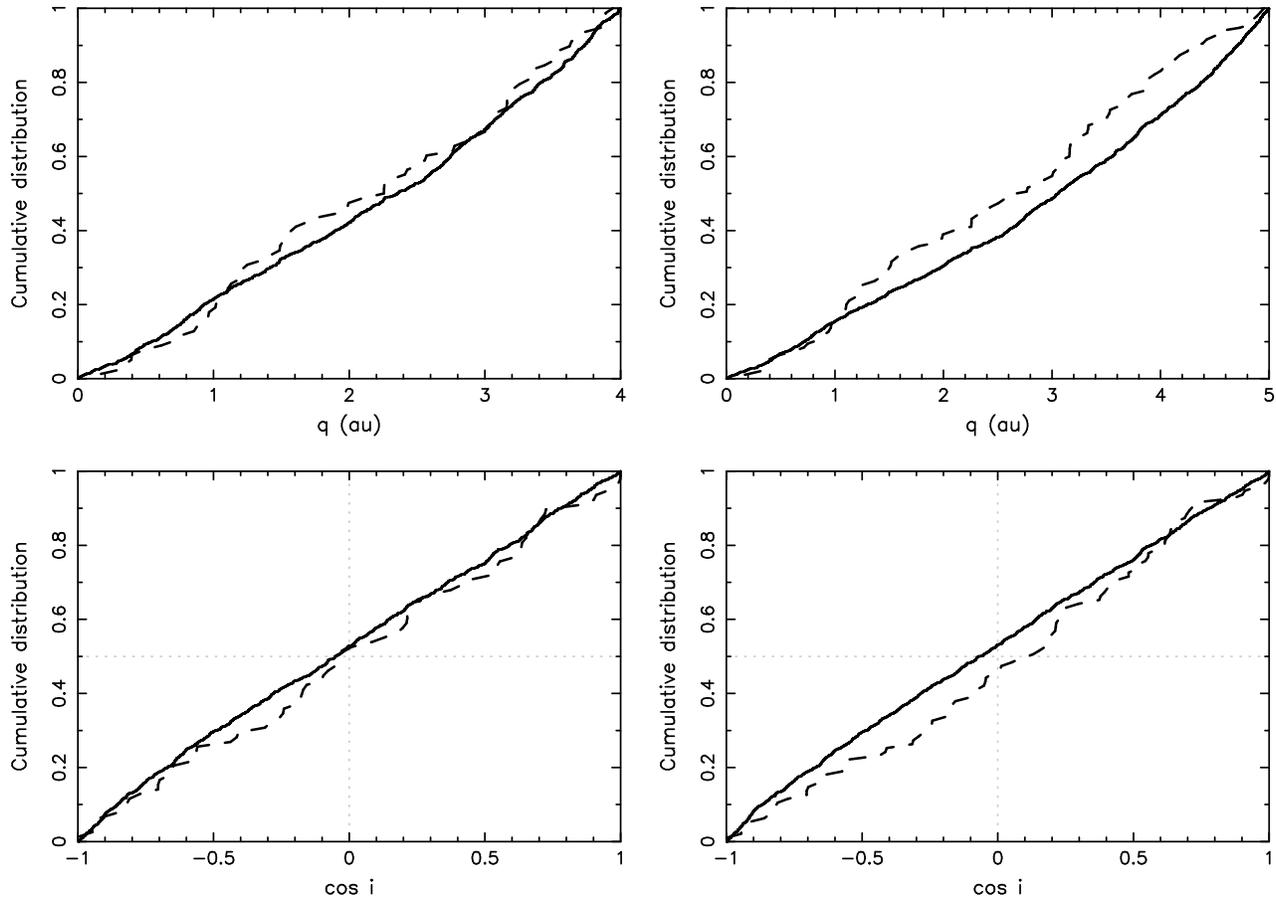}
\caption{Comparison between data (dashed curve) and simulations (solid
 curve) for LPCs on nearly parabolic orbits: cumulative distribution
 of the perihelion distance $q$ (top panels) and cosine of
 inclination $\cos i$ with respect to the ecliptic plane (bottom
 panels). Panels on the left are for a population of LPCs with
 perihelia $q\leq 4$~au, while panels on the right are for a
 population of LPCs with perihelia $q\leq 5$~au. Simulations use our
 {\tt C1V1} run and also include returning comets with Whipple's
 power-law parametrization of the fading law and exponent
 $\kappa=0.6$. Both data and simulated orbits assume $1/a\leq 2\times
 10^{-4}$ au$^{-1}$, i.e., $a > 5000$~au (Fig.~\ref{w_a_fit}) to
 correspond to the nearly parabolic class.}
 \label{w_qi_fit}
\end{figure*}

\subsection{Comparison with observations}\label{res_obs}
We now compare our simulations with the available data summarized in
Sec.~\ref{obs}. We first consider comets on nearly parabolic orbits,
i.e., those in the traditional Oort peak, and then continue with
discussion of all LPCs.

\subsubsection{Nearly parabolic comets}\label{npc}
Our reference data set for this class of orbits comes from the
Kr\'olikowska et~al.\ catalogs described in Section~\ref{war}. We
selected $134$ high-quality orbital solutions with uncertainty in
$1/a$ smaller than $10^{-5}$ au$^{-1}$.  However, the perihelion
distribution of this sample includes few orbits beyond $\simeq 6$~au,
or even less (Fig.~\ref{w_qi}). Therefore, in order to minimize this
bias, we restrict ourselves to a sub-sample corresponding to a smaller
perihelion cutoff. To see the sensitivity to the cutoff limit, we
chose two values: (i) $q\leq 4$~au, and (ii) $q\leq 5$~au, the first
being a more conservative choice.  In what follows, we keep using data
from our simulation {\tt C1V1}, but we checked that the other
simulations produce basically identical results. When handling the
orbits from the last Gyr of the simulations (Section~\ref{out}), we
avoided the $4$~Myr intervals following the strongest stellar
encounters. This prevents confusion with periods of cometary showers.

Figure~\ref{w_a_fit} shows data for the distribution of the original
semimajor axis using the appropriate binding energy $1/a$ instead of
$a$.  The upper panels are for the heliocentric target zone of
$r=4$~au, while the bottom panels assume $r=5$~au. The cometary orbits
contributing to the Oort peak are often approximated with a population
of new comets only \citep[e.g.,][]{fetalAA2017}. Therefore, the left
panels on Fig.~\ref{w_a_fit} use only our simulated LPCs when they
first appear in the target zone. However, Fig.~\ref{nc_ret} showed
that orbits of some returning comets may also occasionally contribute
to the population of nearly parabolic orbits. For that reason, in the
right panels of Fig.~\ref{w_a_fit} we show results from a more
complete, and also more realistic, model where returning comets were
added. We used the Whipple fading power-law model with exponent
$\kappa=0.6$. This was found to be the best value in \citet{WT99} and
also below in Sec.~\ref{rc}. As expected, the modification is not
dramatic when the returning component is added (because most of the
returning orbits have $1/a\geq 2\times 10^{-4}$ au$^{-1}$, not shown
in this figure), but it does improve the comparison with the
data. Each time we normalized the total
simulated population of LPCs to the number of observed data in the
same range of $1/a$ and having the appropriate perihelion cutoff.
This leaves us with $78$ comets for $q\leq 4$~au and $95$ comets for
$q\leq 5$~au, representing $58$\% and $71$\% of the total sample (see
Fig.~\ref{w_qi}). Finally, we divided the simulated orbits into two
classes: (i) those which appeared in the target region when first
recorded on an LPC orbit in our simulations (these are the jumpers),
and (ii) those whose perihelia were recorded to evolve in our
simulation before they entered the target zone (these are the
creepers). Recall that examples of jumpers are in the top panels of
Fig.~\ref{c_orbits}, while examples of creepers are in the bottom
panels. The distribution of $1/a$ values for jumpers is shown with a
blue histogram, while the creepers are shown with a green histogram in
Fig.~\ref{w_a_fit}. For both cutoff values, jumpers represent about
23\% of the whole population. Jumpers represent the old view of how 
Oort Cloud comets became observable, but most observed comets originate 
as creepers. This is in accord with results in
\citet{fetalAA2017}, and previous studies of this group, and with
analyses of previous orbits of directly observed comets
\citep[e.g.,][]{DK2015,KD2017}. As expected, jumpers arrive from the
outermost part of the Oort cloud, for which $a\gtrsim 30000$~au.

Assuming the few hyperbolic orbits among the observed comets are
either interstellar or, rather, solutions where the nongravitational
effects have not yet been accurately modeled, the comparison between
data and model is encouraging. There are two small points of mismatch
both stemming from the fact that the width of the simulated Oort peak 
is slightly smaller than the width of the Oort peak of the observed 
comets \citep[a similar problem has been reported in][]{fetalAA2017}. 
Apparently, a small fraction of the observed comets in the Oort peak 
have too large, or too small, values of the original semimajor axis. 
There are several possible reasons for this problem.

Recall that the original semimajor axes of the observed comets are not
a simple and direct product of the observations. Rather, they have to
be determined by fitting the observations and propagating the orbit
backward in time. This requires that a particular dynamical model be
used. Especially for orbits with nearly zero binding energy to the
Solar system, details play an important role. So, some of the most
extreme orbits of the observed comets in the Oort peak may still have
unrecognized systematic errors. On the other hand, changes in the
parameters of our model might bring better agreement with the data.
For instance, the outer edge of the modeled Oort cloud depends on the
galactic tidal model and especially the assumed value of the local
mass density $\rho_0$. We used $\rho_0=0.15$ M$_\odot$ pc$^{-3}$, but
if the value was smaller, at least during the first $0.5-1$~Gyr of
Solar system evolution, the simulated outer edge of the Oort cloud
would expand. This may, for instance, happen if the Sun was further
from the center of the Galaxy or when solar vertical oscillations with
respect to the galactic plane are included in our model. Our model
also did not include non-gravitational perturbations in the cometary
dynamics. Their absence may explain why our simulated Oort peak is too
narrow on the side of small semimajor axes.

Figure~\ref{w_qi_fit} shows a comparison between the data, observed
comets, and the model for perihelia and inclinations. Here we show the
cumulative distributions of the respective elements and the
simulations include the returning population of comets with the fading
model as above. Only data for comets with nearly parabolic orbits are
used, namely $1/a\leq 2\times 10^{-4}$ au$^{-1}$.  In the panels on
the left side we used a $q\leq 4$~au cutoff, while in the panel on the
right side we used a $q\leq 5$~au cutoff. In the case of our smaller
cutoff, the comparison is again rather satisfactory. While still
noisy, due to the smaller amount of data, the observed distribution of
cometary perihelia is slightly nonlinear, as the model predicts. The
orbital planes are basically isotropic in space. The model predicts a
slight preference for retrograde orbits, in accord with results in
\citet{fetalAA2017}. The data do not provide clear evidence for this
effect, perhaps due to the still small sample of comets.

When extending the target zone to $5$~au (right-side panels in
Fig.~\ref{w_qi_fit}), the match between data and our model becomes
worse. This is especially seen in the distribution of perihelia. It is
likely that the observations still missed some comets with
perihelia beyond $4$~au, and those on retrograde orbits, but this
issue can only be resolved with more data from future surveys.
\begin{figure*}[t!]
\epsscale{1.1}
\plotone{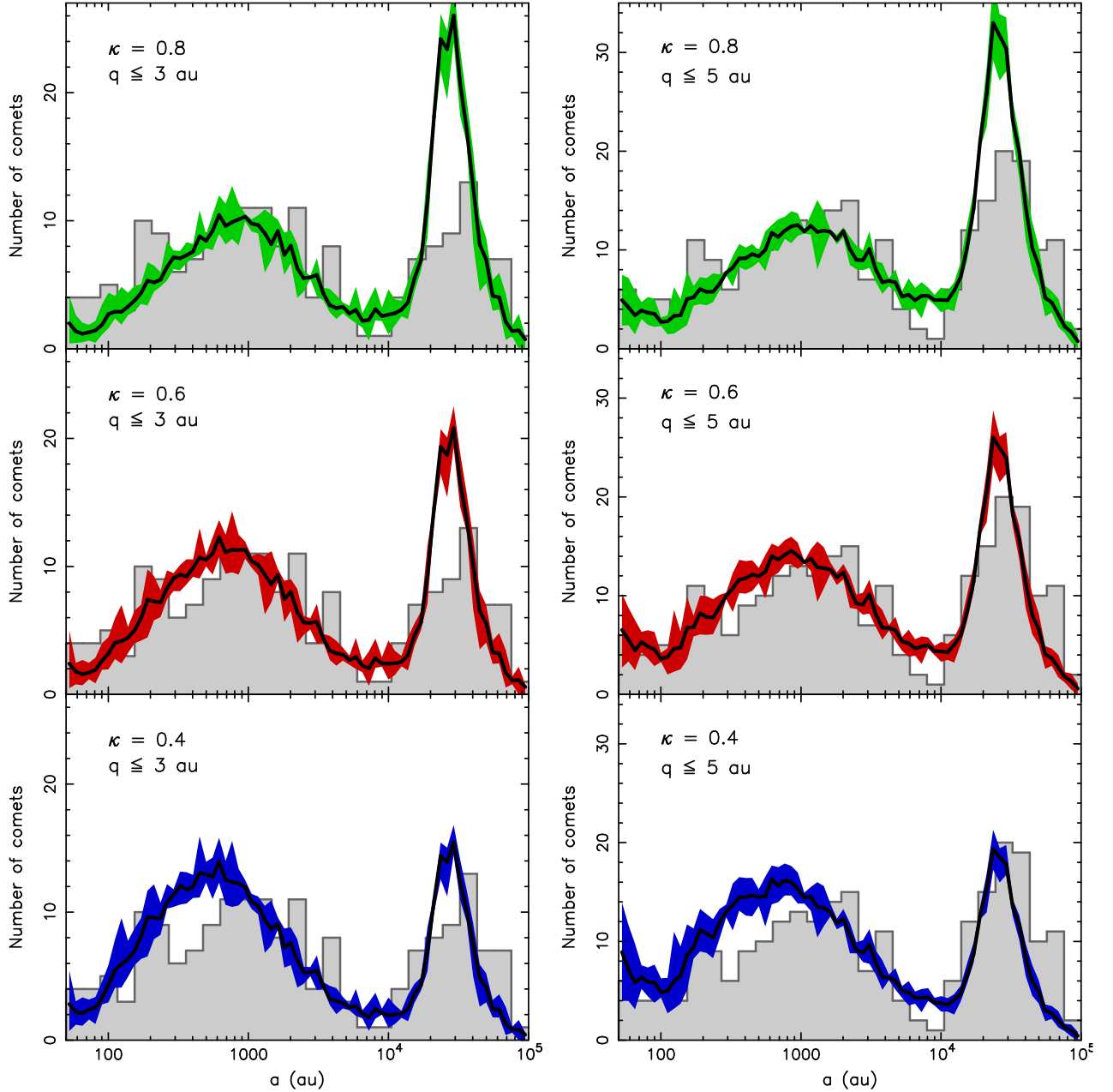}
\caption{Comparison between data (gray histogram) and simulations for
 the whole population of LPCs: incremental distribution of the
 original semimajor axes $a$.  Data use class 1 solutions in the
 MWC08 catalog with two constraints on perihelia: (i) left panels for
 $q\leq 3$~au (191 orbits), and (ii) right panels for $q\leq 5$~au
 (267 orbits). Simulations assume Whipple's fading law with three
 different values of the power-law exponent $\kappa=0.8$ (top),
 $\kappa=0.6$ (middle), and $\kappa=0.4$ (bottom). Results from all
 four of our simulations (see Table~\ref{tab1}) are used: the black
 line is their mean value and the color region is delimited by the
 minimum and maximum value from the runs. The simulated distributions
 are normalized to the total number of data points.}
 \label{mw_a_fit}
\end{figure*}
\begin{figure*}[t!]
\epsscale{1.1}
\plotone{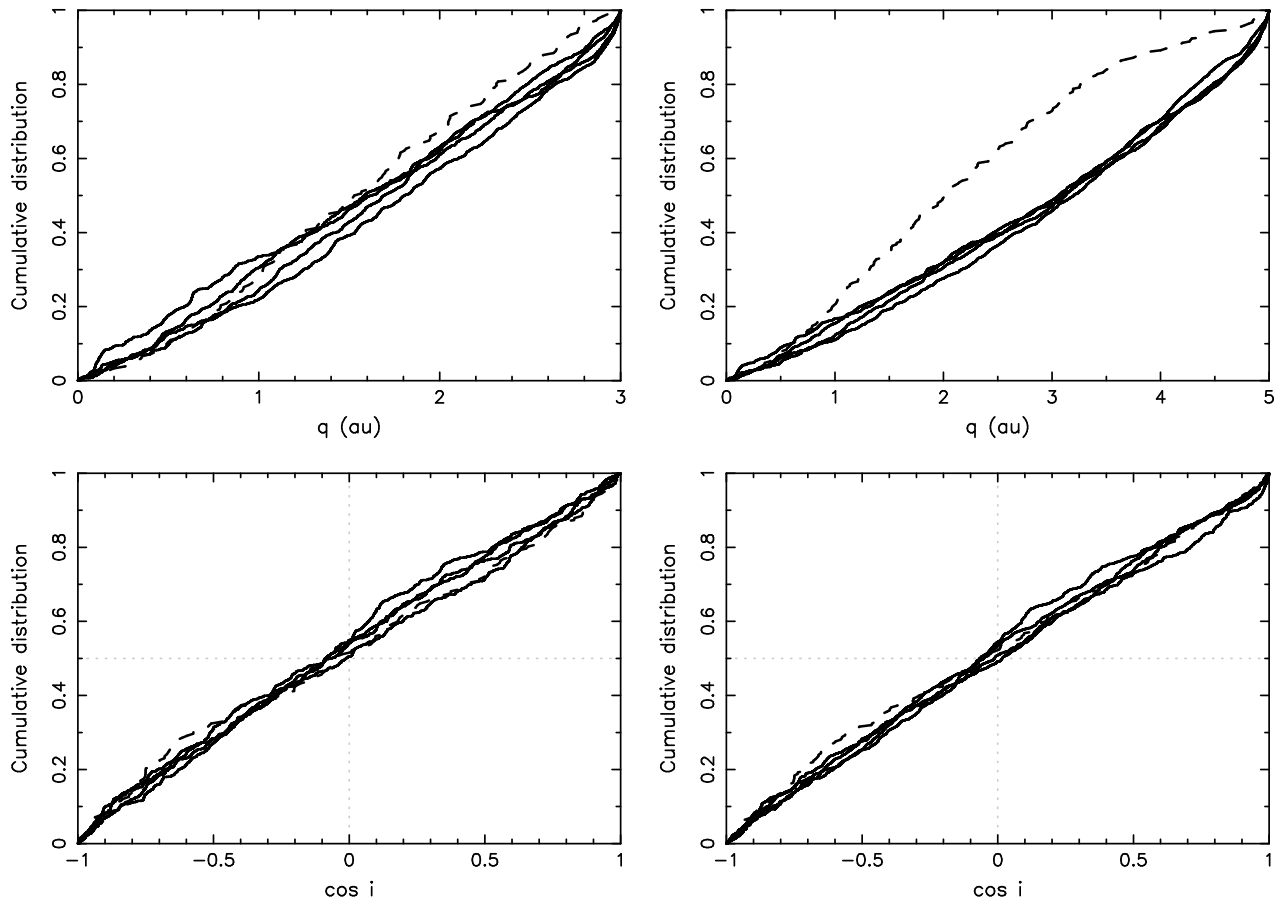}
\caption{Comparison between data (dashed line) and simulations (solid
 lines) for the whole population of LPCs: cumulative distributions of
 perihelia (top) and cosine of ecliptic inclination (bottom). Data use
 class~1 solutions in the MWC08 catalog: (i) left panels for $q\leq 3$~au
 orbits, and (ii) right panels for $q\leq 5$~au orbits. Solid lines are
 results from our four simulations (see Table~\ref{tab1}).}
 \label{mw_qi_fit}
\end{figure*}

\subsubsection{All long-period comets}\label{rc}
As outlined in Section~\ref{mwc}, the reference source for the orbits
of all LPCs, including the returning ones, is the MWC08 catalog. It
contains 318 accurate orbits (classes 1A and 1B), whose distribution
of original semimajor axes was shown in Fig.~\ref{mw_a}. However, the
distribution of their perihelia (Fig.~\ref{mw_qi}) suggests the sample
is still fairly incomplete beyond $q\simeq 4$~au. In order to minimize
the influence of this bias, we consider sub-samples of the whole MWC08
catalog by setting a limit on the perihelion distance. For the sake of
comparison, we consider two cases: (i) $q\leq 3$~au, likely less
biased, but containing a smaller number of observed comets, and (ii)
$q\leq 5$~au, a larger sample, but already seeing the onset of
bias. As in the previous Section, we avoid the periods of the
strongest comet showers in the output from our simulations in their
last Gyr. To see the variance of the results of our four jobs (see
Table~\ref{tab1}), we now use all these runs. We compute the mean
value of the parameter of interest and report its minimum and maximum
values among the four jobs. In all cases we use Whipple's fading law
described in Section~\ref{fad} with the only free parameter $\kappa$
being the power exponent of the life expectancy $\Phi_n$ through the
$n$-th perihelion return. A larger value of $\kappa$ corresponds to a
faster fading of new comets, while a smaller value of $\kappa$
emphasizes the role of the returning population of comets.

Figure~\ref{mw_a_fit} illustrates the match between the MWC08 data and
our suite of simulations for comets with $q\leq 3$~au (left panels)
and $q\leq 5$~au (right panels). Each time, we show results for three
values of the $\kappa$ exponent: (i) $\kappa=0.8$ (top), (ii)
$\kappa=0.6$ (middle), and (iii) $\kappa=0.4$ (bottom). Note that
\citet{WT99} obtained $\kappa=0.6\pm 0.1$ from their
analysis. Clearly, the choice $\kappa=0.8$ unsuitably increases the
signal in the Oort peak over the continuum of returning LPCs with
semimajor axes $\leq 10000$~au for either choice of the perihelion
cutoff.  For the $5$~au perihelion limit, the middle value
$\kappa=0.6$ appears to do the best job, and decreasing $\kappa$ to
$0.4$ would already give too much weight to the population of the
returning comets if compared to the Oort peak comets. Restricting the
perihelion limit to $3$~au only, even $\kappa=0.4$ provides an
acceptable result. So values $\kappa$ in the range $0.4$ to $0.6$ seem
promising. In fact, assuming that the comet incompleteness beyond
perihelia $\simeq 4$~au is dominated by the lack of observed returning
comets, $\kappa=0.4$ may also satisfy the observations if some
returning orbits are added.

We tried to complement such a qualitative analysis with a more
rigorous, quantitative treatment of fitting the model predictions to
the data.  We used the MWC08 observations distributed in 27 equal-size
bins in $\log a$, as shown in Fig.~\ref{mw_a_fit}. We formally assumed
$\sqrt{N}$ uncertainty statistics. We then ran the traditional
least-squares fit of the fading law exponent $\kappa$ using results
from our simulations. After performing this effort, we indeed obtained
best values of this formal $\chi^2$ at $\kappa\simeq 0.6$ for $q\leq
5$~au orbits, but we noticed that the minimum normalized $\chi^2$
value was larger than unity (between $1.2$ and $1.4$, depending on our
simulation). Restricting the orbits to $q\leq 3$~au, the formal best
fit value shifted to $\kappa\simeq 0.5$, and the minimum normalized
$\chi^2$ values were between $1.15$ and $1.25$.  At face value, this
should imply rejection of the model. We admit that the model is
imperfect in many aspects. First, the determination of cometary orbits
may have its problems, but perhaps more importantly, the fading model
may be just too simple. On the other hand, we also believe that the
data still suffer unrecognized systematic errors and
incompleteness. As we are not able to remove these issues with the
available data set, the least-squares model is plainly a formal
procedure that confirms the qualitative analysis from above, but
cannot improve it in a more objective way. If anything, the formal
$\chi^2$ suggests $\kappa=0.6^{+0.1}_{-0.2}$ is the best fit to the data. 
This is the same result as \citet{WT99} obtained, although we have a 
slightly larger error bar. Our allowance for slightly smaller values of 
$\kappa$ follows from our feeling that the population of returning 
comets is observationally underrepresented.

Figure~\ref{mw_qi_fit} shows a comparison between data and simulations
for the cumulative distributions of perihelion and cosine of orbital
inclination with respect to the ecliptic. The left panels are for the
cutoff limit $q\leq 3$~au, while the right panels are for the cutoff
limit $q\leq 5$~au. In this case, we use only the simulation with
$\kappa=0.6$, since inspection of other choices shows these results
are not sensitive to the $\kappa$ value. The conservative restriction
on perihelia, $q\leq 3$~au, leads to a fairly satisfactory match
between data and model predictions. This is not surprising. Even if
the suspected systematic errors bias the original values of the
semimajor axes in the MWC08 catalog, the values of perihelia and
inclination are much less dependent on the model uncertainty.
Incompleteness may play some role beyond $q\simeq 2$~au, as may be
suggested by the top left panel in Fig.~\ref{mw_qi_fit}. Obviously,
things get much worse when the looser cutoff limit of $q\leq 5$~au is
adopted. Here, the lack of observed comets with perihelia beyond
$q\simeq 2-3$~au is obvious. In all cases, though, the isotropy of the
orbital planes in space seems to match the available data.
\begin{figure}[t!]
\epsscale{1.1}
\plotone{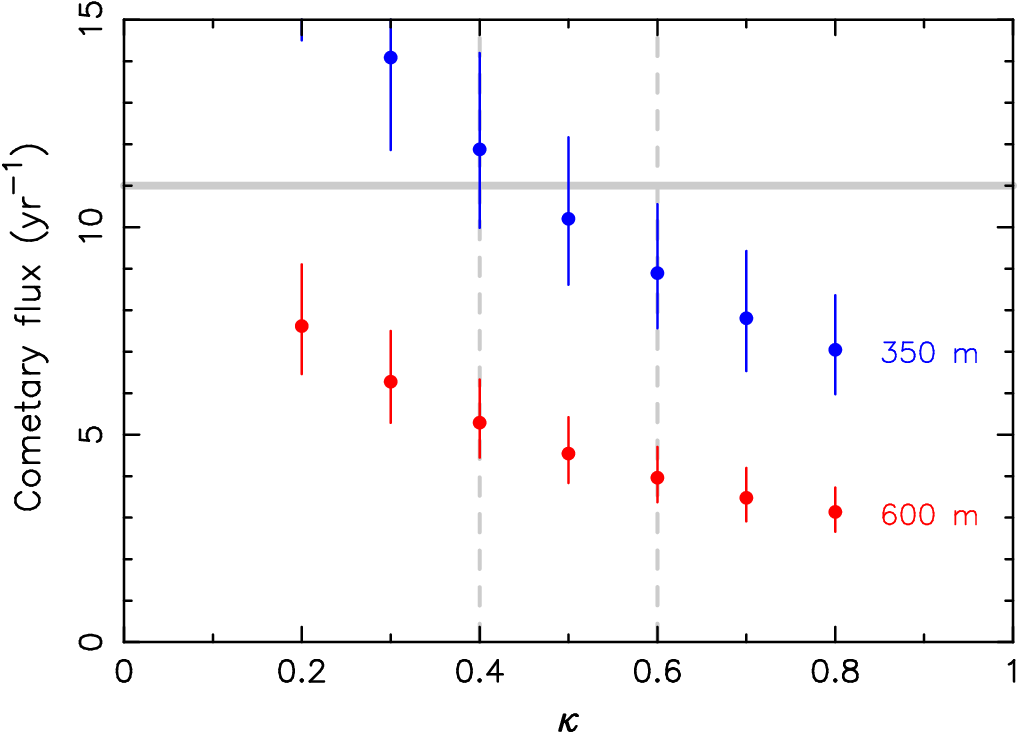}
\caption{Predicted annual flux of LPCs with $q\leq 4$~au (ordinate) as
 a function of the power-law exponent $\kappa$ of Whipple's
 fading law (abscissa).  Results from all our four simulations (see
 Table~\ref{tab1}) are used: circles give their mean value and the
 interval indicates minimum to maximum values in the runs. Output
 from the simulations was calibrated for two size values of comets:
 (i) $D=600$~m in red, and (ii) $D=350$~m in blue.  The horizontal
 gray line, $11$ comets per year, is the flux for $H\leq 10.9$ LPCs
 reported by \citet{francis2005}. The interval of $\kappa$ values
 between $0.4$ and $0.6$, vertical gray dashed lines, is our
 preferred range from analysis of the original semimajor axis
 distribution.}
 \label{flux}
\end{figure}

\subsubsection{Flux of long-period comets}\label{fl}
Finally, we confront the observed flux of LPCs with predictions from
our model. We remind the reader that an obstacle to an exact
comparison is that (i) the observed flux is magnitude-limited, while
(ii) our model predictions are size-limited. The trouble arises
because the size vs.\ magnitude relation for comets is uncertain.

We first focus on the model predictions. Previous applications of our
framework allowed us to calibrate the trans-Neptunian planetesimal
disk, namely the population of particles constituting the initial
conditions of our simulations.  There were about $8\times 10^{11}$
planetesimals with $D\geq 1$~km in this region, a value that may be
about $50$\% uncertain \citep{nesvornySPC}. The cumulative size
distribution between $1$ and nearly $100$~km may be approximated by a
power law with an exponent $\simeq -2$, while the cumulative size
distribution below $1$~km is uncertain, but may follow a power law
with a shallower exponent $\simeq -1.5$ \citep[see Fig.~14
  in][]{nesvornySPC}. Note that the extrapolation to sub-kilometer
sizes is not as well-constrained as the size distribution for bodies
with diameters $\geq 1$~km.

Next, we consider the population of simulated LPCs in the last Gyr of
our runs whose orbits have a certain perihelion cutoff. In what
follows, we take $q\leq 4$~au. Running analysis with different
power-law exponents $\kappa$ of the Whipple fading scheme, we
determine how many comets our simulations predict. We again avoid the
$4$~Myr periods following the strongest stellar encounters. Having this
information, we can readily predict an unbiased number of LPCs with a
given size $D$ reaching, on average, their perihelia each
year. Choosing two sizes $D=350$~m and $D=600$~m, we obtained the
fluxes shown in Fig.~\ref{flux}: symbols give the mean value over our
four simulations, while the associated interval indicates the minimum
and maximum fluxes from these jobs. The predicted flux is correlated
with $\kappa$: smaller values of this parameter lead to larger fluxes,
and vice versa. This result makes sense, because a smaller $\kappa$
value lets comets live longer by surviving more perihelion
returns. Our preferred values $\kappa\simeq 0.4-0.6$ imply a flux of
$\simeq 3-6$ LPCs per year with $D \geq 600$~m and $q < 4$~au. Note that the
indicated range of the fluxes formally follows from predictions given
by our four runs. In each case, we used a mean value of comet flux
over a long time span of 1~Gyr. Additionally, the flux fluctuates about 
this mean value by up to $25$\%; see the upper panels in Figs.~\ref{nc_20},
\ref{nc_5} and \ref{nc_ret}. Data in those figures used large bins in
time, $4$~Myr, but these simulations also had a limited number of
comets integrated, $10^6$ initially. Assuming the product of these two
parameters is roughly constant, the estimated fluctuations apply to
kilometer-sized comets over time period of a decade or few.

Having the fading law calibrated by the majority of observed LPCs, we
also predict that the largest comet observable in two centuries should
have a diameter of $\simeq (16-20)$~km. However, this is certainly an
underestimate, because large comets fade much less than small
comets. Consider, for instance, that HTCs, which evolve from LPCs,
have been shown to typically fade only after about $3000-5000$ returns
\citep[e.g.,][]{nesvornySPC}. Therefore, the flux of large comets like
Hale-Bopp (C/1995 O1) is underestimated by our analysis.  For the sake
of a test, we completely disregarded fading and analyzed the
statistics of the observed LPCs. From this we would predict the
largest LPCs seen over two centuries should have a size between $32$
and $38$~km, still somewhat smaller than the estimated sizes of the
largest LPCs such as Hale-Bopp 
(see review in \citet{fer2002}; cf.\ \citet{Hui2018}). However,
we deal with the statistics of a few objects, which may be subject
to larger fluctuations.

On the side of the observed population of LPCs we recall that
\citet{francis2005} estimated a flux of about $11$ LPCs with $q\leq
4$~au and $H\leq 10.9$ annually.  Assuming the magnitude-size relation
from \citet{sf2011}, this magnitude limit would correspond to a size
of about $600$~m. (Note, however, that the magnitude-size relations
used by \citet{francis2005}) imply larger nuclei, with diameters
$\approx 1-2$~km; see Section~\ref{flu}.) Data in Fig.~\ref{flux} show
that our model prediction falls short of predicting this flux by a
factor of about two or three. In order to align the results with
observations, the common LPCs
should have a size of $\simeq 350$~m (see the blue symbols). Given the
hyperactivity of LPCs, this may not be unreasonable. To check whether
the flux-prediction problem could be caused by the simplicity of the
fading law we used, we also computed the annual flux of new comets
with $q\leq 5$~au. Assuming a typical size $D=350$~m, we obtained
$4.1\pm 0.9$ (sampling again results from our four simulations). This
would favorably compare with the stated $4$ new comets in this region
annually \citep[e.g.][]{fetalAA2017}. Therefore, the fading law is
likely not a problem for the flux determination.  If, however, the
comet flux should be higher, as indicated by the analysis of
\citet{betal2017}, or the magnitude vs.\ size relation should require
a larger size than assumed here, the model prediction would be below
the observed population of LPCs.

We thus find once again that modeling the LPC flux is the most
problematic issue of their analysis. Most often researchers infer the
Oort cloud population from the LPC flux. This is obviously a circular
argument as far as the predictive power is concerned.
Whenever previous studies attempted to use independent calibrations of
the Oort cloud population, the estimates ran short of explaining the
LPC flux. For instance, \citet{bm2013} considered model-predicted
constraints on the ratio between the populations of the scattering
disk and the Oort cloud. To reconcile the observed fluxes of
Jupiter-family comets (JFCs) and LPCs, \citet{bm2013} concluded that
LPCs must be systematically smaller at the same absolute magnitude
than JFCs. Nonetheless, their remaining mismatch was still a bit
larger than our factor of $\simeq 2-3$. Until nuclear sizes of LPCs are
accurately determined from observations, most likely from future
surveys of comets at large heliocentric distances, we are left with a
couple of speculations.

Either LPCs are typically small, as suggested above, or the Oort cloud
population is larger than we obtained. One possibility is that our
assumed population of kilometer-sized planetesimals in the original
trans-Neptunian disk was underestimated. In fact, their number is not
constrained directly by any of the implantation processes into reservoirs
of small bodies (such as Jupiter Trojans), but is
set by the assumption of a
shallow size distribution of the disk particles at small sizes. This
is suggested by the paucity of small craters on Pluto and Charon,
compared with a collisional distribution \citep[e.g.,][]{rob17, Singer2019}. If
those craters formed, but were then erased, however, the initial size
distribution of small planetesimals might be steeper than we assumed.
Another possibility is to deliver more comets into the Oort cloud than
expected from our model. This could happen during the early phase when
the Solar system was still in the birth cluster, or by considering a
larger planetesimal source zone than we did here. Recall that our
initial disk was limited to the region from Neptune's orbit to about
$30$~au. If planetesimals on initial orbits that are closer to or
further from the Sun can also contribute, the Oort cloud population
might be somewhat larger. Analysis of these possibilities is left for
future studies.
\begin{figure*}[t!]
\epsscale{1.1}
\plotone{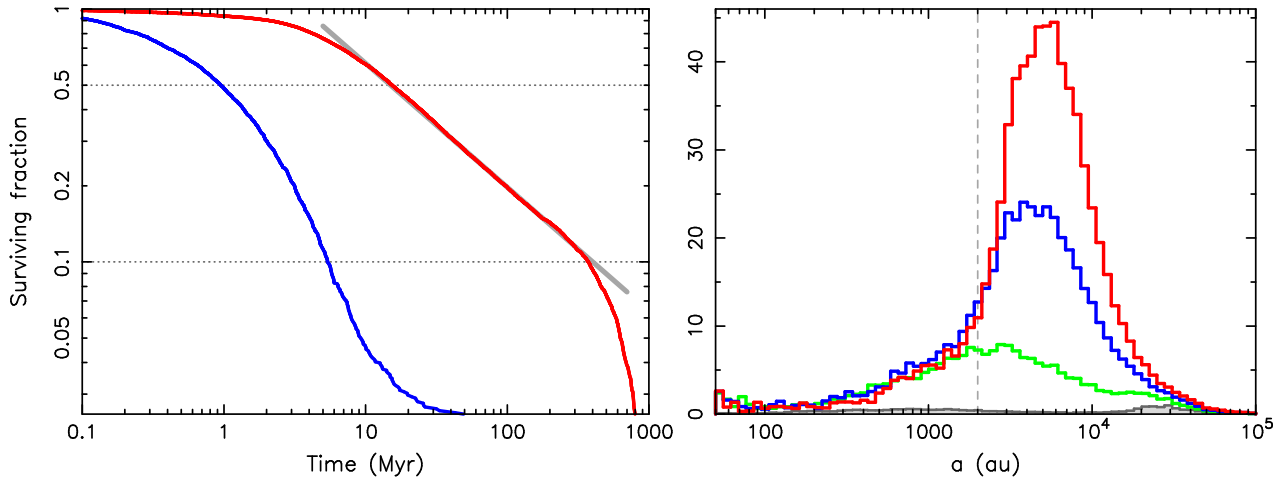}
\caption{Left panel: Distribution of the number of comets that remained in
 the monitored target zone longer than time $t$. Data from the last
 Gyr from all our simulations combined are used, and two heliocentric
 radii $r$ of the target zone are plotted: (i) $r=5$~au (blue curve),
 and (ii) $r=20$~au (red curve). No fading is imposed, so particles
 are only eliminated dynamically. The gray line shows the $\propto
 t^{-1/2}$ power-law which matches the $r=20$~au data between $10$
 and $200$~Myr well.  Right panel: The same as in
 Fig.~\ref{nc_ahist}, but now with all returning particles. The
 distribution of original semimajor axis values for all comets in the
 last Gyr of our simulation {\tt C1V1} whose perihelia are in
 heliocentric target zones with four radii are shown: (i) $r=5$~au
 (gray histogram), (ii) $r=10$~au (green line), (iii) $r=15$~au (blue
 line), and (iv) $r=20$~au (red line). Cometary fading is applied
 only when the perihelion gets smaller than $5$~au (Whipple power-law
 scheme with $\kappa=0.6$). The ordinate is arbitrarily normalized to
 the maximum of the $r=5$~au plot. The total number of comets in
 $r\leq 20/15/10$~au zones is $\simeq 29.2/19.9/9.0$ times larger
 than the population entering the $r\leq 5$~au zone. The dashed
 vertical line points out the edge of the inner Oort cloud in our
 simulation.}
 \label{lsst}
\end{figure*}


\section{Predictions for future surveys}\label{future}
Looking ahead to the future, perhaps the most interesting result in
this paper is the prediction of a significant increase of the LPC
population beyond perihelion distance $\simeq 15$~au
(Fig.~\ref{nc_qdist}). This is not a shocking conclusion. It has
already been discussed in some previous studies
\citep[e.g.,][]{st2016,fetalAA2017}, which were, however, based on
simpler dynamical models. Here we present the most important features
as they are predicted by our simulations. These results are mainly
relevant for future well-characterized surveys such as with the Large
Synoptic Survey Telescope (LSST). Unlike \citet{st2016}, who
implemented basic features of a magnitude- and time-limited survey
with a specific sky coverage, we present only the model, unbiased
prediction. Constraints imposed by biases of a specific survey are not
considered here.

\citet{st2016} noted that LPCs with distant perihelia can roam in the
trans-Saturnian region for a very long time. We find the same
result. The red line in the left panel of Fig.~\ref{lsst} shows the
distribution of time spent by comets on orbits with $q\leq 20$~au
during the last Gyr of our simulations (we combined results from all
runs).  From about $10$~Myr, the surviving fraction of comets in this
wide target zone falls-off only very slowly and is well approximated
with a $t^{-1/2}$ power-law. The longest survival times, beyond about
$600$~Myr, are missing in our data, but this is a result of the
restricted time interval in which we monitor the cometary orbits. Very
likely, the tail of the distribution will reach beyond a billion
years. The exponent $-1/2$ of the time dependence is characteristic of
a random walk of the orbital perihelia beyond $15$~au, at a safe
distance from both Saturn and Jupiter 
\citep[also see][]{yab79,st2016}. At the same time, the inner part of the Oort
cloud can inject LPCs onto these distant-perihelion orbits with small
semimajor axes, thus strongly gravitationally bound to the Solar
system.  Galactic tides and tugs from passing stars then feed the
random walk diffusion with small steps persisting for a very long
period of time.  In contrast, LPCs whose perihelia reach the currently
observable zone with $q\leq 5$~au, say, have the dynamical
survivibility distribution shown by a blue line in the left panel of
Fig.~\ref{lsst}. These times are much shorter, a result of the
typically larger semimajor axes of these comets. They are thus more
weakly bound to the planetary system and are at high risk of being
ejected by perturbations due to the gas giants (either direct or
indirect, reflected in the motion of the Solar system barycenter).

The above-described surviving fraction distribution uses all visits of
LPCs into the specified target zone within the last Gyr of our
simulations. If we were to ask how much time the population of LPCs
observed within the last century, say, have already spent wandering in
the $q\leq 20$~au zone, the distribution would still be skewed to
longer times. This is because only a fraction of short times (say
$\leq 10$~Myr) would be relevant to that task as we now fix the position
of the interval in time. We find
that more than $50$\% of LPCs with distant perihelia observed ``now''
have already spent more than $100$~Myr cruising the target zone, and
some $15$\% were injected into the planet-crossing zone more than half
a billion years ago. This is in accord with conclusions in
\citet{st2016}.

The right panel in Fig.~\ref{lsst} shows similar information as
Fig.~\ref{nc_ahist}, but now extended by a population of returning
comets.  We use the results from the {\tt C1V1} simulation and plot a
distribution of original semimajor axis values for the predicted
steady-state population of LPCs reaching heliocentric zones with
different perihelion cutoffs of $5/10/15/20$~au. Following methods in
\citet{st2016}, we assume fading for comets with small perihelia only,
in our case $q\leq 5$~au, certainly an approximation which needs to be
refined in future comparison of the model predictions and the
observations. The barely-seen gray histogram near the bottom of the
plot corresponds to the currently observed population of LPCs with
perihelia $q\leq 5$~au; this distribution is identical to that in the
middle right panel in Fig.~\ref{mw_a_fit}. The color-coded lines correspond 
to LPCs in larger perihelia zones, namely $10$~au (green), $15$~au (blue), 
and $20$~au (red). Increasing the cutoff limit has two main implications:
(i) the number of LPCs increases rapidly beyond $15$~au (e.g., it is
nearly $30$ times larger for $q\leq 20$~au, compared to the $q\leq
5$~au population); (ii) for large $q$ cutoffs the semimajor axis peaks
in the inner Oort cloud zone, as this region can now efficiently feed
these orbits. Our simulations show a steep drop of the distributions
at about $1500$~au (blue and red lines), which reflects the edge of
the created Oort cloud (see Fig.~\ref{oort_aei}). However, should this
edge prove to be closer -- for instance, due to the existence of a
fossilized inner Oort cloud extension from the birth-cluster phase of
Solar system evolution -- the distributions shown in the right panel
of Fig.~\ref{lsst} would also extend to smaller $a$ values. Here
again, only comparison of the model predictions with the observations
will help to solve this issue.

We find that the modeled population of LPCs with perihelia $q\leq
5$~au has a slight preference for retrograde orbits (also see
Fig.~\ref{mw_qi_fit}). This is in accord with predictions from other
models, such as \citet{fetalAA2017} and \citet{st2016}. However, the
population with the largest perihelia in our model, say between
$15$~au and $20$~au, shows a preference for prograde orbits
(representing about $65$\% of the whole sample in this category). This
conclusion differs from that in \citet{st2016}. Recall, however, that
\citet{st2016} assumed an isotropic extension of the Oort cloud to its
innermost part. As shown in Sec.~\ref{our_cloud}, this assumption is
not correct. The inner Oort cloud below semimajor axis $\simeq
7000$~au is strongly anisotropic, reflecting its origin in the
scattering disk.  Comets arriving from this part of the Oort cloud,
which is the majority among the distant-perihelia orbits, remember the
anisotropy of their source zone in our model.

In order to make our model useful, we prepared software which exports
our results in the form of an unbiased population simulator (codes and
results are available from the authors upon request). Choosing a
heliocentric zone $r\leq 20$~au, it allows the user to create a
catalog of LPC orbits with perihelia $q\leq r$ and whose orbits are
statistically compatible with the orbital distribution from our
simulations, assuming a steady-state situation. Our model also
provides the LPC flux for bodies of different sizes following from the
assumed initial population in the trans-Neptunian, comet-birth
disk. Users can load the catalog and apply the observability
efficiency of a specific survey. This way, the unbiased set of orbits
from our model can generate a specific set of observable comets which
can be compared with the data.


\section{Conclusions}\label{concl}
With the advent of new all-sky surveys in the forthcoming decade,
we constructed a numerical model describing the origin and orbital
evolution of comets. The strength of our approach consists of
its being a unified scheme for all comets, both short- and long-period.  
The short-period comet part has been described
at length in \citet{nesvornySPC}. Here we dealt with the
long-period comets.

The model has several aspects. Its primary justification comes from
confrontation with observations. To that end we collected all
currently available data about LPCs. Surprisingly, the orbital
distribution of the observed comets can still be reasonably well
matched with only minimal tuning. The principal phenomenon we solved
for is the LPC fading law. With the limited range of perihelion
distances for which the observed sample is reasonably complete, the
single-parameter model of \citet{whipple1962} is
sufficient. The remaining differences between the data and the model
predictions are small and they are plausibly explained by persisting
observational biases.  That said, certainly the model may also be
improved in a number of aspects, but without understanding the data
better, we do not see a strong need to make the model more
complex. As to the LPC flux, the comparison between the observations
and model is less good. While several model simplifications may be
responsible for these differences, we believe that they are mainly
because of the poorly-understood relationship between the size of a
cometary nucleus
and its absolute brightness. The model uses the sizes, while the
observations provide the magnitudes. Attempts to
link the two are still not completely satisfactory. Again, until 
these problems are resolved, far-reaching modifications of the model
seem not to be justified.

Things, however, will change soon when powerful upcoming all-sky
surveys will start providing observations. As far as LPCs are
concerned, the crucial aspect is the extension of the perihelion range
to at least $15-20$~au. Such data will offer a much more complete
mapping of the Oort cloud, the source region of the long-period
comets. This is because current observations effectively sample
only the outermost isotropic tail of this vast source population. Its
critical inner zone, still hidden to our data, contains much more
information diagnostic of the history of the Solar system (both as far
as its natal conditions and also the giant planets' late
migration). These future LPC observations will be able to 
directly probe the whole Oort cloud. Because of the likely much
smaller activity of LPCs at large heliocentric distances, these new
observations will also help us to clarify the current uncertainties
related to their flux.

Our model allows us to provide a useful first glimpse of the expected
number of comets with distant perihelia. At this moment, however, we
do not feel safe to turn them into specific quantitative predictions
for two reasons. First, we do not have complete information about
complex observational biases, such as magnitude limits, exposure
times, sky-coverage cadence, etc. In this situation, it makes more
sense to provide an unbiased population prediction and work
iteratively with a specific survey to fine-tune the model parameters
by comparing its predictions with observations. We completed this
task, but consider it a zero-order attempt. This is because an unknown
aspect, likely also to be inferred from the observations, is the
activity and fading of LPCs at large perihelion distances. Therefore,
more advanced versions of the LPC population prediction need to be
completed in the future.


\acknowledgements
We thank the referee, Mike Kelley, for several comments that helped to improve
the final version of this paper, and Gerbs Bauer, Piotr Dybczy{\'n}ski, 
Matthew Knight, Ma\l{}gorzata Kr{\'o}likowska, Hal Levison, and Maik Meyer
for discussions. This research was supported by the Czech
Science Foundation (grant 18-06083S). D.N.'s work was supported by the
NASA Emerging Worlds program.

\end{document}